\documentclass[12pt]{article}
\usepackage{amssymb,amsmath}
\usepackage{young}
\usepackage{graphicx}
\usepackage{ulem}
\usepackage{cite}
\begin{document} 
\begin{center} 
{\bf\Large {A study of nuclear radii and neutron skin thickness of neutron-rich 
nuclei near the neutron drip line} }
\end{center} 
\vspace{.1in} 
\begin{center} 

{\bf Usuf Rahaman,M.Ikram, M. Imran, Anisul Ain Usmani}\\
Department of Physics, Aligarh Muslim University, Aligarh-202002, India\\

\end{center} 
\begin{center} 
{\bf Abstract} 
\end{center} 
We studied the charge radius ($r_{c}$), neutron radius ($r_n$), 
and neutron skin-thickness ($\Delta r=r_n-r_p$) 
over a chain of isotopes from C to Zr with the stable region to the neutron drip line. 
Theoretical calculations are done with axially deformed self-consistent relativistic 
mean-field theory (RMF) with effective nonlinear NL3 and NL3* interactions. 
The theoretically estimated values are compared with 
available experimental data and a reasonable agreement is noted.
We additionally assessed the two-neutron separation energy ($S_{2n}$) 
to mark the drip line nuclei of the considered isotopic series. 
In the reference of $S_{2n}$, neutron magicity is also discussed.
The calculated neutron radii are compared with empirical estimation made by $r=r_0N^{1/3}$ 
to examine the abnormal trend of the radius for neutron drip line nuclei.
In view to guide the long tails, the density distribution for some skin candidates is analyzed. 
Finally, neutron skin thickness is observed for the whole considered isotopic series.  
\vskip 1 cm
{\bf PACS}: 21.10.k, 21.10.Dr, 21.10.Ft, 21.10.Gv, 21.60.n
\vskip .5 cm
{\bf Keywords}: Relativistic mean-field (RMF) theory; Skin thickness; Neutron-rich nuclei.

\newpage

\section{Introduction}
The study of exotic neutron-rich nuclei has gained worldwide attention
because of their unexpected behaviour such as nuclear halo and skin~\cite{Tanihata,horo1,horo2}.
These phenomena are supposed to be significant characteristics of the nuclei at 
the extreme~\cite{horo1,horo2}. The pieces of information gathered from 
such nuclei are used for astrophysical studies
to understand the origin of heavy elements. The
exotic nuclei having excessive number of neutrons possess
larger $N/Z$ ratios than the $\beta$ stable nuclei.
These excessive neutrons in neutron-rich nuclei are 
responsible to produce a huge contrast in
Fermi energy of neutrons and protons, in turn, 
making a decoupling between neutron and 
proton distribution and as a result, nuclear skin structure is formed.

In past few years, a large number of theoretical studies have been conducted 
to determine the nature of the neutron distribution in exotic neutron-rich 
nuclei~\cite{hagen2016,kaufmann2020,Agbemava2014,xia2018,zhang2020}.
Agbemava {\it et al.} has made recent progress on neutron drip line 
nuclei within the RMF model 
using several parameterizations in axially manner~\cite{Agbemava2014}. 
The investigations have been used 
for global assessment of the accuracy of the description of 
nuclear structure phenomena especially 
ground-state properties of even-even nuclei and also to describe the 
location of two-proton as well as two-neutron drip lines.
Moreover, Xia {\it et al.} also calculated the ground-state properties of 
nuclei with $8\le Z\le 120$ using the spherical relativistic continuum Hartree-Bogoliubov method~\cite{xia2018}.
They described that there are 9035 nuclei predicted to be bound by incorporating 
the continuum effect which largely extends the existing nuclear landscapes.
They demonstrated that the coupling between the bound states and the continuum due 
to the pairing correlations plays an essential role in extending the nuclear landscape.
In continuation of the above work, K. Zhang {\it et al.}, made the description of even-even 
nuclei in the nuclear chart in an axially deformed manner 
based on point coupling method~\cite{zhang2020}.
In Ref.~\cite{xia2018}, the work is done for a wide range of nuclear chart within spherical 
symmetry but the present investigations have to be made for axially 
symmetric cases whereas a point 
coupling method is used in Ref.~\cite{zhang2020}. 
Moreover, present work is also devoted to all even-odd, odd-even, odd-odd, and even-even systems.
Therefore, it has significance in nuclear structure phenomena 
especially the nuclear skin structure of exotic nuclei.

It merits referencing that the density distributions of exotic neutron-rich nuclei 
are quite different from the nuclei reside at the stability line. 
The profile of neutron density is supposed to be extended beyond 
the proton density as the excessive neutrons are pushed out against the nuclear surface and 
therefore creating a sort of neutron skin. 
The neutron skin thickness is characterized as $\Delta r = r_n - r_p$ with $r_n$ and 
$r_p$ being the root mean square (RMS) radii for the neutron and proton distributions, respectively.
The development of neutron skin on the surface of a nucleus is a marvel of enthusiasm for nuclear 
structure physics to represents the rudimental nuclear properties of nuclei. 
The neutron skin thickness depends mainly on the balancing condition between the isospin 
asymmetry and the Coulomb force. 
Thus, the skin is responsible enough to look at the isovector properties while our understanding of isovector density 
($\rho_1=\rho_n-\rho_p$) is poorly known. 
However, in the mean-field estimations, the skin thickness is identified with 
the divergence in the Fermi energies among protons and neutrons.
The main perception on neutron skin with a thickness of 0.9 fm was accounted 
for by Tanihata {\it et al.}~\cite{Tanihata1992} in He nucleus, and later was 
additionally affirmed by G. D.Alkhazov {\it et al.}~\cite{Alkhazov1997}.

Reliable theoretical calculations regarding neutron skin are 
quite essential not only to describe the structure of 
neutron-rich nuclei but also for modeling the neutron-rich matter.
In recent years, a large number of theoretical studies have been conducted to determine the 
nature of the neutron distribution in exotic neutron-rich nuclei~\cite{Brown2017,Brown2009,Centelles2017,Min2006}.
The relativistic calculations on neutron skin thickness have been made for the $^{208}$Pb nucleus and 
its relationship with the slope of symmetry energy has been examined~\cite{warda,chen,Brown2000,Horowitz2001,Typel2001,Furnstahl2002,Karataglidis2002}.
Apart from theoretical investigations, the recent Lead Radius EXperiment (PREX) 
has now established the existence
of a neutron skin for the $^{208}$Pb nucleus in a clean and 
model-independent way with a high level of accuracy~\cite{prex}.
Moreover, new experimental programs using both stable and exotic beams at various 
laboratories such as JLab, FAIR, FRIB, MESA, and RIKEN are in progress and the data are awaited concerned to 
the nuclear structure studies in order to find the answers of some fundamental questions over 
the neutron skin thickness~\cite{thiel2019}.

Neutron skin thickness of neutron-rich nuclei is intently related to density dependence 
of symmetry energy and a significant discernible for testing 
the symmetry potential of nuclear matter~\cite{horo1,horo2}.
Neutron skin plays a significant role in correlation with several physical observable over 
the finite nuclei, nuclear symmetry energy and infinite 
nuclear matter to pure neutron 
matter~\cite{chen,Brown2000,Horowitz2001,Typel2001,Furnstahl2002,
Karataglidis2002,centelles,Agarwal,Mondal,Reinhard,Angeli}. 
The determination of the neutron skin thickness for 
finite nuclei is supposed to be a 
unique experimental constraint on the symmetry energy. 
Skin varies linearly with the slope 
parameter $(L)$ of density dependence of the  nuclear 
symmetry energy at saturation density 
and it can probe the isovector part of the nuclear interaction~\cite{chen}.
Moreover, it is also beautifully linked with the various constraint 
found to force on the neutron equation of state (EOS) of high-density matter in neutron 
stars~\cite{Brown2000,Horowitz2001,Typel2001,Furnstahl2002,Karataglidis2002,Agarwal}.
The EOS has a significant impact on neutron star structure modeling~\cite{Horowitz2001,steiner,carriere,ban,meng}.
Despite many efforts in infinite many-body systems, our knowledge of density dependence of the symmetry 
energy is still very limited~\cite{chen,Brown2000,Horowitz2001,Typel2001,Furnstahl2002,Karataglidis2002}.
Nuclear symmetry energy can not be measured directly within the available nuclear experimental 
facilities but the information about it tends to be  picked up by the assurance of either the neutron 
skin of neutron-rich nuclei or the radii of neutron 
stars~\cite{chen,Brown2000,Horowitz2001,Typel2001,Furnstahl2002,Karataglidis2002}. 
In this way, predictions on neutron skin might be useful to fix the constraint on the calculation 
of symmetry energy of infinite nuclear matter or pure neutron matter which in turn can be used 
to simulate the mass and radius of the neutron star.
In this regard, we endorse a measurement of the neutron skin thickness of the neutron-rich nuclei from C to Zr isotopes.

In this paper, we make a theoretical investigation for neutron skin for a series of 
neutron-rich nuclei from C to Zr isotopes. 
The present calculations are performed within self-consistent axially deformed relativistic 
mean-field model with effective NL3 and NL3*  parameter sets. The physical observable of interest are 
root mean square charge radius, neutron radius, neutron skin thickness, separation energy and density 
distributions of protons and neutrons. 
In this work, the drip line signifies the  two-neutron drip line calculated by two neutron separation energy.
Moreover, the exposure of the closed shell of the  nuclei are also discussed on 
the basis of two-neutron separation energy. 
The paper is arranged in the following way: Section one contains the introduction of the manuscript. 
The used formalism relativistic mean-field model is expressed in section two. Results are given in section three. 
Finally, the manuscript is summarized and concluded in section four.

\section{Theoretical Formalism}
The RMF theory has made incredible progress in portraying the nuclear many-body problem and 
also explained the numerous nuclear phenomena over the whole periodic table 
~\cite{S92,GRT90,R96,SW86,BB77,yadav,skpatra,Lalazissis1998}. 
It is very better to get the spin orbit splitting automatically 
over the non-relativistic case which gains us to understand 
the closed shell structure of the nuclei~\cite{sharma,warrier}.
The RMF theory starts with the fundamental Lagrangian 
density containing nucleons interacting 
with $\sigma-$, $\omega-$, and $\rho-$meson fields. The photon field $A_{\mu}$ is 
incorporated to deal with the Coulomb interaction of protons. 
The Lagrangian density for the relativistic mean-field theory is represented as~\cite{S92,GRT90,R96,SW86,BB77},

\begin{eqnarray}
{\cal L}&=&\bar{\psi_{i}}\{i\gamma^{\mu}
\partial_{\mu}-M\}\psi_{i}
+{\frac12}\partial^{\mu}\sigma\partial_{\mu}\sigma
-{\frac12}m_{\sigma}^{2}\sigma^{2}
-{\frac13}g_{2}\sigma^{3} \nonumber \\
&-&{\frac14}g_{3}\sigma^{4}-g_{s}\bar{\psi_{i}}\psi_{i}\sigma 
-{\frac14}\Omega^{\mu\nu}\Omega_{\mu\nu}+{\frac12}m_{w}^{2}V^{\mu}V_{\mu}\nonumber \\
&-&g_{w}\bar\psi_{i}\gamma^{\mu}\psi_{i}
V_{\mu}-{\frac14}\vec{B}^{\mu\nu}\vec{B}_{\mu\nu} 
+{\frac12}m_{\rho}^{2}{\vec{R}^{\mu}}{\vec{R}_{\mu}}-{\frac14}F^{\mu\nu}F_{\mu\nu} \nonumber\\
&-&g_{\rho}\bar\psi_{i}\gamma^{\mu}\vec{\tau}\psi_{i}\vec{R^{\mu}}-e\bar\psi_{i}
\gamma^{\mu}\frac{\left(1-\tau_{3i}\right)}{2}\psi_{i}A_{\mu} .
\end{eqnarray}
Here M, $m_{\sigma}$, $m_{\omega}$, and $m_{\rho}$ are the masses for nucleons, 
${\sigma}$-, ${\omega}$-, and ${\rho}$-mesons, and ${\psi}$ is its Dirac spinor. 
The field for the ${\sigma}$-meson is denoted by ${\sigma}$, ${\omega}$-meson 
by $V_{\mu}$ and ${\rho}$-meson by $R_{\mu}$. 
$g_s$, $g_{\omega}$, $g_{\rho}$ and $e^2/4{\pi}=1/137$ are the coupling 
constants for the ${\sigma}$-, ${\omega}$-, ${\rho}$-mesons, and 
photon respectively.$g_2$ and $g_3$ are the self-interaction coupling constants for
 ${\sigma}$-mesons. By utilizing the classical variational principle,
 we get the field equations for the nucleons and mesons. 
The Dirac equation for the nucleons is inscribed by
\begin{equation}
\{-i\alpha\bigtriangledown + V(r_{\perp},z)+\beta M^\dagger\}\psi_i=\epsilon_i\psi_i.
\end{equation}
The effective mass of the nucleons is
\begin{equation}
M^\dagger=M+S(r_{\perp},z)=M+g_\sigma\sigma^0(r_{\perp},z),
\end{equation}
and the vector potential is
\begin{equation}
 V(r_{\perp},z)=g_{\omega}V^{0}(r_{\perp},z)+g_{\rho}\tau_{3}R^{0}(r_{\perp},z)+
e\frac{(1-\tau_3)}{2}A^0(r_{\perp},z). 
\end{equation}
The field equations for mesons are given by 
\begin{eqnarray}
\{-\bigtriangleup+m^2_\sigma\}\sigma^0(r_{\perp},z)&=&-g_\sigma\rho_s(r_{\perp},z)\nonumber\\
&-& g_2\sigma^2(r_{\perp},z)-g_3\sigma^3(r_{\perp},z) ,\\
\{-\bigtriangleup+m^2_\omega\}V^0(r_{\perp},z)&=&g_{\omega}\rho_v(r_{\perp},z) ,\\
\{-\bigtriangleup+m^2_\rho\}R^0(r_{\perp},z)&=&g_{\rho}\rho_3(r_{\perp},z) , \\
-\bigtriangleup A^0(r_{\perp},z)&=&e\rho_c(r_{\perp},z). 
\end{eqnarray}
The meson field ${\sigma^0}$, $V^0$, $R^0$ are only the time dependent component of the meson field. 
These components come into play when time reversal symmetries are taken into consideration for solving the field equations.
Under time reversal, the spatial components of meson fields are omitted.
As we know that relativistic quantities have four components; 1-time and 3-spatial ($i=0,1,2,3$). 
As the time reversal symmetry is used, the spatial component has been eliminated and therefore 
only the time component is used to play which is denoted by $i=0$ and therefore meson fields are denoted by ${\sigma^0}$, $V^0$, $R^0$.

Here, $\rho_s(r_{\perp},z)$, and $\rho_v(r_{\perp},z)$ are 
the scalar and vector density for $\sigma$- and $\omega$-fields 
in a nuclear system and represented as
\begin{eqnarray} 
\rho_s(r_{\perp},z) &=& \sum_ {i=n,p}\bar\psi_i(r)\psi_i(r)\;,                         \nonumber \\
\rho_v(r_{\perp},z) &=& \sum_{i=n,p}\psi^\dag_i(r)\psi_i(r) \;.			      
\end{eqnarray}

These equations of motion are solved to acquire a static solution for the nuclei to deduce their 
ground state properties.
The set of nonlinear coupled equations are solved self-consistently 
in an axially deformed harmonic oscillator basis $N_F=N_B=14$ for fermion and boson basis. 
The radii are calculated from the corresponding densities
\begin{eqnarray}
\langle r_p^2\rangle &=& \frac{1}{Z}\int r^{2}d^{3}r\rho_p(r_{\perp},z)\;,        \nonumber \\
\langle r_n^2\rangle &=& \frac{1}{N}\int r^{2}d^{3}r\rho_n(r_{\perp},z)\;,        \nonumber \\
\langle r_m^2\rangle &=& \frac{1}{A}\int r^{2}d^{3}r\rho(r_{\perp},z)\;,          
\end{eqnarray}
for proton, neutron, and matter rms radii, respectively.
The quantities $\rho_p$, $\rho_n$, and $\rho$ are their 
relating densities. 
The charge rms radius can be calculated from the proton rms radius 
utilizing the standard expression~\cite{ekstrom2015,hu2016} 
\begin{equation}
\langle r_{c}^2 \rangle =\langle r_p^2 \rangle+R_p^2+\frac{N}{Z} R_n^2+\frac{3h^2}{4m_p^2c^2}, 
\end{equation}
where  $3h^2/4m_p^2c^2 \simeq 0.033$fm$^2$, $R_n^2=-0.1149(27)$fm$^2$, $R_p=0.8775(51)$fm.

The quadrupole deformation parameter $\beta_{2}$ is extracted from
 the calculated quadrupole moments of neutrons and protons through 
\begin{equation}
Q = Q_n + Q_p = \sqrt{\frac{16\pi}5} \left(\frac3{4\pi} AR^2\beta_2\right),
\end{equation}
where $R=1.2A^{1/3}$.\\
The total energy of the system is given by 
\begin{equation}
 E_{total} = E_{part}+E_{\sigma}+E_{\omega}+E_{\rho}+E_{c}+E_{pair}+E_{c.m.},
\end{equation}
where $E_{part}$ is the sum of the single particle energies of the nucleons and 
$E_{\sigma}$, $E_{\omega}$, $E_{\rho}$, $E_{c}$, $E_{pair}$, $E_{c.m.}$ are 
the contributions of the meson fields, the Coulomb field, pairing energy,  
and the center-of-mass energy, respectively. 

In present calculations, the non-linear NL3~\cite{lalazissis1997} and 
NL3*~\cite{lalazissis2009} parameterizations are used throughout the calculations. 
In Table ~\ref{tab1}, we have listed the masses, coupling constants
for nucleons and mesons for both NL3 and NL3* effective force parameterization sets. 
The properties of nuclear matter for the same parameterizations are also framed in the Table ~\ref{tab1}.
For detailed formalism and numerical techniques, it is recommended to follow the
Refs.~\cite{gambhir,patra} and the references therein. 
It is known that consideration of pairing is important for open shell nuclei.
Thus, constant gap BCS approximation is used to take care of 
pairing interaction as given in Ref.~\cite{madland}.
In the case of simple BCS prescription, the 
expression of pairing energy is written by 
\begin{equation} 
E_{pair}=-G\bigg[\sum_{i\;>0}u_iv_i\bigg]^2\;,
\end{equation}
where G is the pairing force constant, and $v_i^2$ and 
$u_i^2=1-v_i^2$ are the occupation probabilities.
The variation with respect to the occupation numbers, $v_i^2$, 
is expressed by the well-known BCS equation
\begin{equation} 
2\epsilon_iu_iv_i-\bigtriangleup(u_i^2-v_i^2)=0\;,
\end{equation}
with $\bigtriangleup=G\sum_{i>0}u_iv_i$.
The occupation number $n_i$ is given by
\begin{equation} 
n_i = v_i^2 =\frac{\displaystyle1}{\displaystyle2}\Bigg[1-\frac{\displaystyle \epsilon_i-
\lambda}{\displaystyle \sqrt{(\epsilon_i-\lambda)^2+\bigtriangleup^2}}\Bigg]\; ,
\end{equation}
where $\epsilon$ is the single-particle energy for the state $i$. 
The chemical potential $\lambda$ for protons (neutrons) is obtained requiring 
\begin{equation} 
\sum_{i} n_i= Z(N).
\end{equation}
The sum is taken over proton (neutron) states. 
The value of constant gap (pairing gap) for proton and neutron are 
determined from the phenomenological formula of Madland 
and Nix~\cite{madland} which are given as 
\begin{equation} 
\bigtriangleup_n = \frac{r}{N^{1/3}}exp(-sI-tI^2)\;, 
\end{equation}
and
\begin{equation} 
\bigtriangleup_p = \frac{r}{Z^{1/3}}exp(-sI-tI^2) \;, 
\end{equation}
where $I=(N-Z)/A$, $r=5.73$ MeV, $s=0.117$, and $t=7.96$.
In particular, for the solution of the RMF equations with 
pairing, we never calculate the pairing force constant G explicitly. 
But the occupation probabilities are directly calculated using the gap 
parameters ($\bigtriangleup_n$ and $\bigtriangleup_p$) 
and the chemical potentials ($\lambda_n$ and $\lambda_p$) for 
neutrons and protons, while chemical potentials are determined 
by the numbers of protons and neutrons. 
And now, the expression of pairing energy is simplified to
\begin{equation} 
E_{pair}=-\bigtriangleup\sum_{i\;>0}u_iv_i \;.
\end{equation}
The centre-of-mass correction is included by non relativistic 
expression i.e. $E_{c.m.}=-\frac{3}{4}41A^{-1/3}$.

Moreover, it is not enough to compute the binding energy and quadrupole moment of odd-$N$ or odd-$Z$
or both odd-$N$ and odd-$Z$ systems in the present model due to mean-field approach.
Therefore, we use the Pauli blocking approximation to take care 
of the time reversal symmetry in the mean-field 
model and pursued our calculations in this context.
The blocking approximation restores the time-reversal symmetry and as a result, reveals 
the even-odd staggering very nicely, but doubles our effort as we need to perform our
calculations twice~\cite{patra2001,usmani2018}.

\begin{table}
\centering
\caption{The NL3~\cite{lalazissis1997} and NL3*~\cite{lalazissis2009} force parameter sets and 
their corresponding nuclear matter properties~\cite{lalazissis1997,lalazissis2009,Dutra,Grill} are listed here.}
\renewcommand{\tabcolsep}{0.3cm}
\renewcommand{\arraystretch}{1.1}
\small
\begin{tabular}{|cc|cc|}
\hline
parameters && NL3 & NL3* \\
\hline
 ${M} $ (MeV)           &   &	939  	&	939	\\
 ${m}_{\sigma} $(MeV)	&	&	508.194 &	502.5742 \\
 ${m}_{\omega} $(MeV)	&	&	782.501	&	782.6 \\
 ${m}_{\rho}   $(MeV) 	&	&	763  	&	763 \\
 ${g}_{\sigma} $      	&	&	10.217	&	10.0944	\\
 ${g}_{\omega} $      	&	&	12.868	&	12.8093	\\
 ${g}_{\rho}  $       	&	&	4.474	&	4.5748	\\
 ${g}_2$  (fm$^{-1}$) 	&	&  -10.4307	& -10.8093	\\
 ${g}_3$              	&	&  -28.8851	& -30.1486	\\

\hline
\multicolumn{4}{|c|}{Nuclear matter properties}                \\
\hline
 $\rho_o$ (fm$^{-3}$)	&(Saturation density)	                    &	0.148	&	0.15	\\
 $(E/A)_{\infty}$ (MeV) &(Infinite nuclear matter binding energy)  	&	-16.29	&	-16.31	\\
 $m^{*}/m$		        &(Ratio of effective mass and bare mass)    &	0.595	&	0.594	\\
 $K$ (MeV)              &(Incompressibility)     	               	&	271.5	&	258.27	\\
J$=$S$(\rho_{0})$ (MeV)	&(Saturation density)	                    &	37.4	&	38.68	\\
$L$ (MeV)               &	(Slope of S)	                        &   118.65	&	122.63	\\
$K_{sym}$ (MeV)       	&(Curvature of S)	                        &	101.34	&	105.56	\\
\hline
\end{tabular}
\label{tab1}
\end{table}

\section{Results and discussions}
In this manuscript, we played out a self-consistent axially 
deformed relativistic mean-field calculation
to find out the binding energy, radii, quadrupole deformation parameter, 
and density distributions. 
Two-neutron separation energy and skin-thickness are estimated 
from the calculated binding energies and radii, respectively.
Both separation energy and neutron radius are supposed 
to be crucial parameters to determine the halo or skin 
structure of neutron-rich exotic nuclei. And, 
thus the determination of nuclear skin within the considered 
isotopic series is the main objective of our manuscript.
Further to guide the view of long tails, the nucleons density distribution is analyzed.
We also discuss the appearance of closed shell within the considered nuclei.
All the results in the forms of tables and figures are explained in forthcoming subsections.


\begin{figure}
\centering
\vspace{0.4cm}
\includegraphics[scale=0.4]{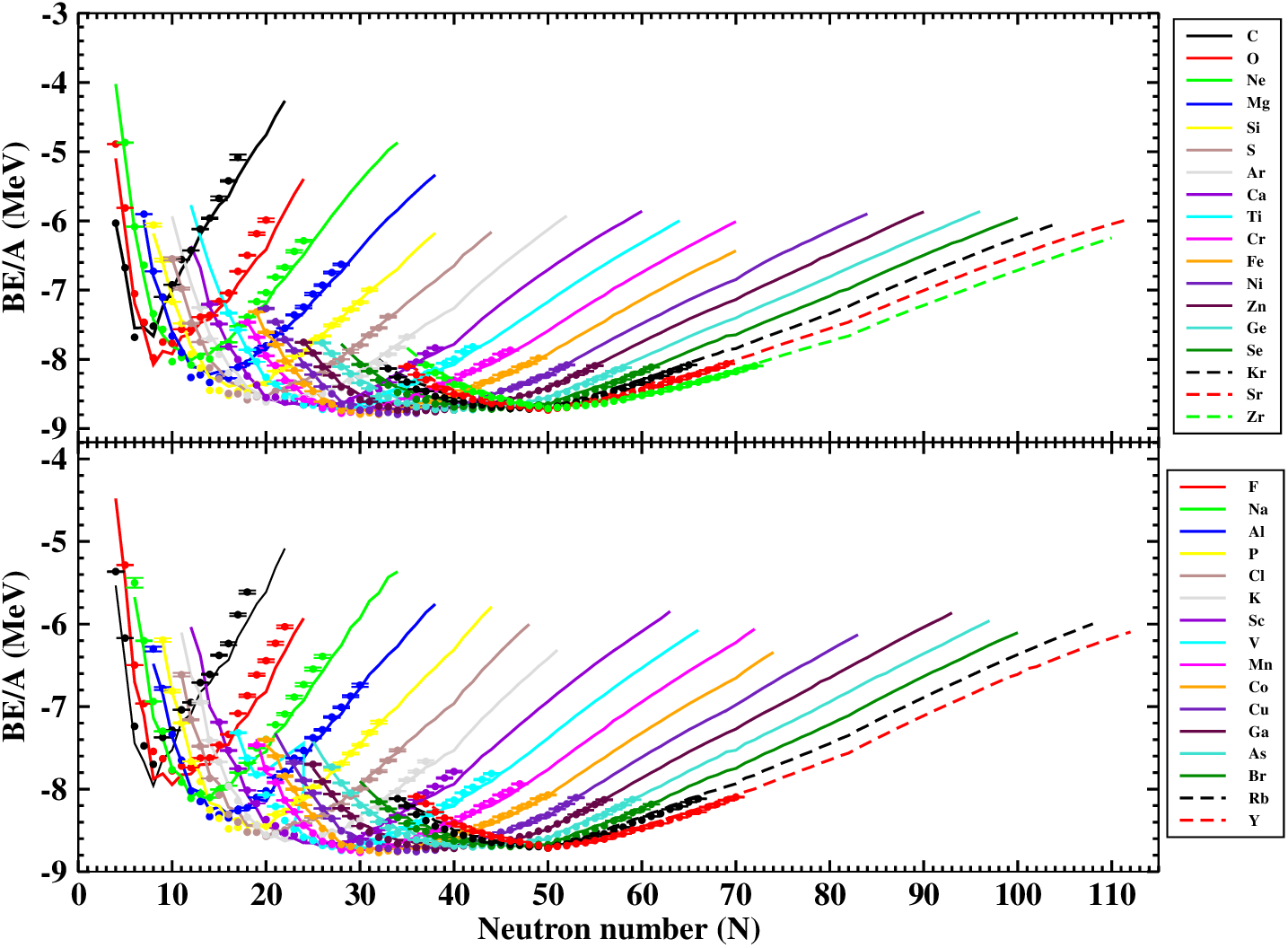}         
\caption{(color online) Binding energy per particle ($BE/A$) calculated using
RMF theory with NL3 parameterization, is 
plotted as a function of neutron number for nuclei from  C to Zr. Experimental data are represented by filled circles.
Even-$Z$ nuclei on the upper panel and odd-$Z$ nuclei are given on the lower panel.}
\label{be1}
\end{figure}

\begin{figure}
\centering
\vspace{0.4cm}              
\includegraphics[scale=0.4]{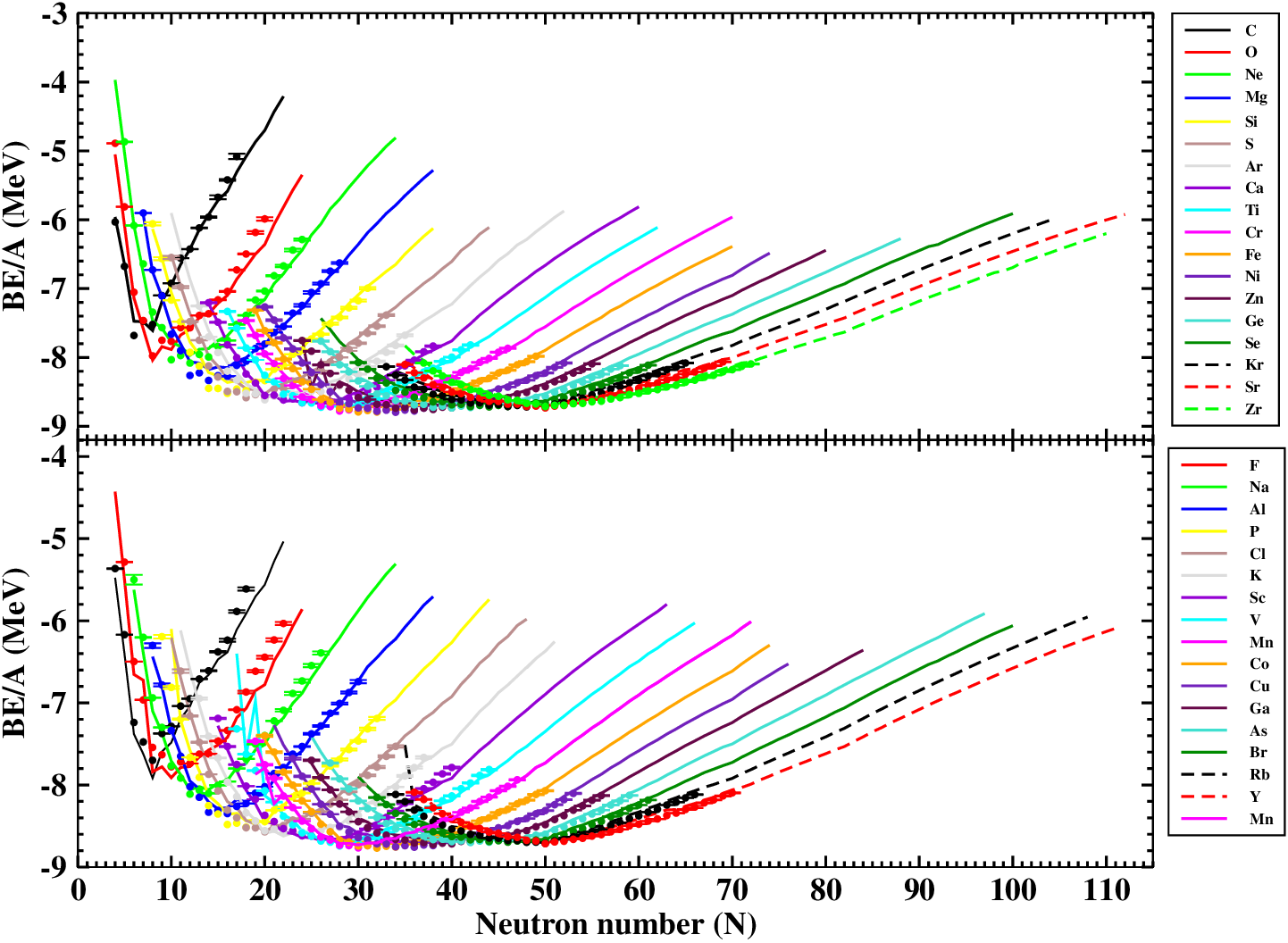}                   
\caption{(color online) same as Fig.~\ref{be1} but for NL3* parameterization.}
\label{be2}
\end{figure}

\begin{figure}
\centering
\vspace{0.4cm}
\includegraphics[scale=0.45]{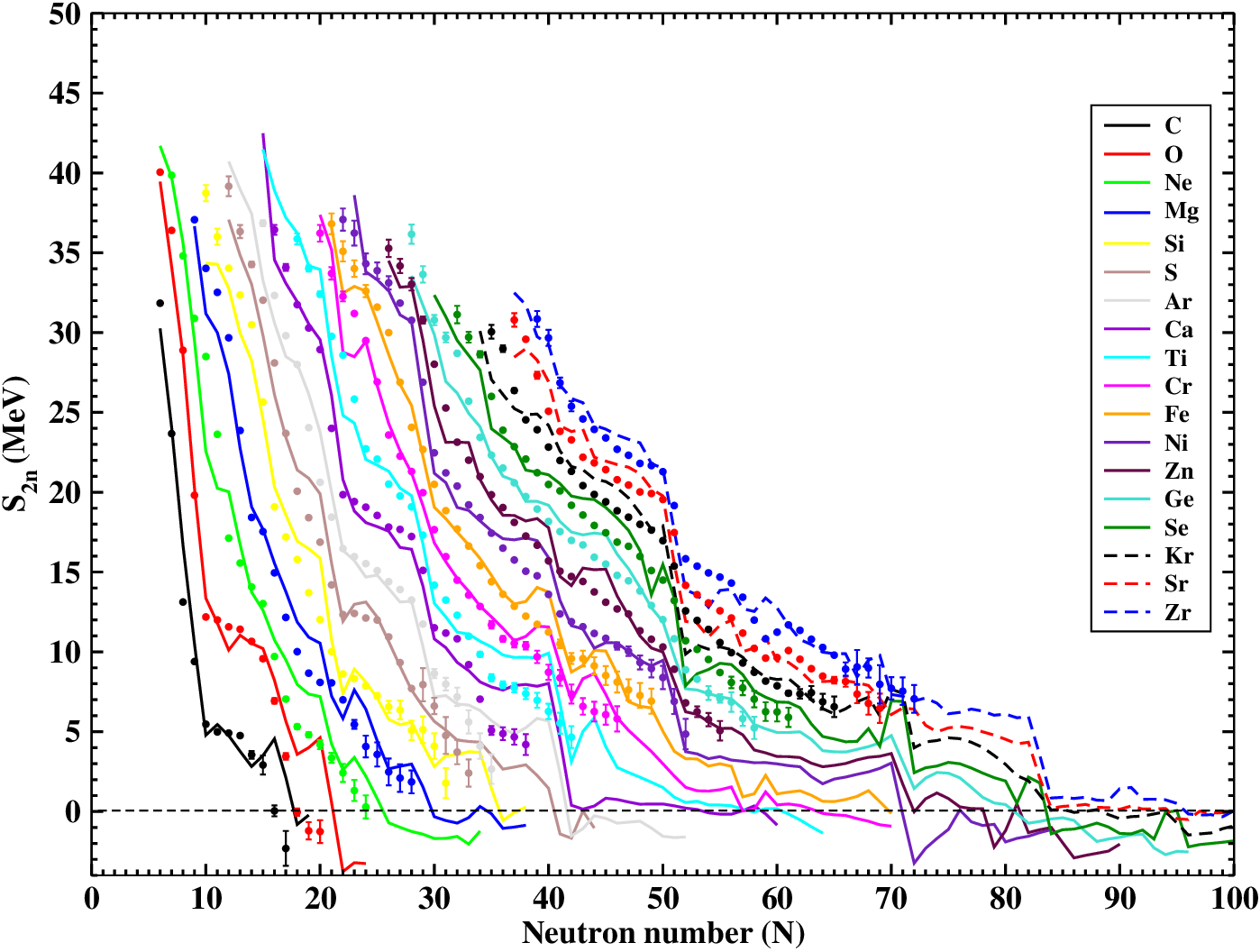}
\caption{(color online) Two neutron-separation energy ($S_{2n}$) calculated using  RMF theory with NL3 parameterization, 
is plotted as a function of neutron number  for even-$Z$ isotopes from C to Zr. Experimental data are represented by filled circles.}
\label{sn1}
\end{figure}

\begin{figure}
\centering
\vspace{0.4cm}
\includegraphics[scale=0.4]{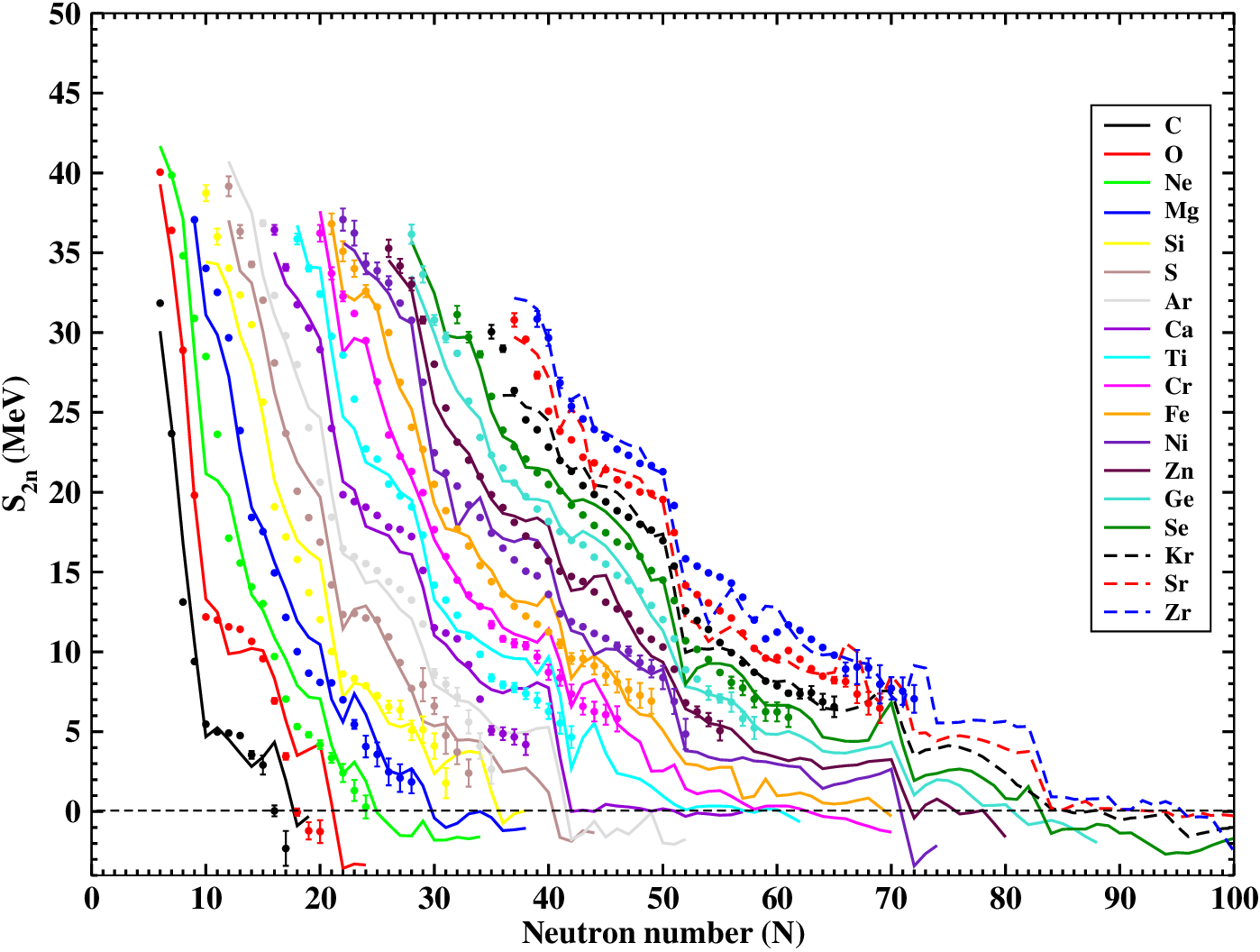}
\caption{(color online) same as Fig.~\ref{sn1} but for NL3* parameterization.}
\label{sn2}
\end{figure}
\begin{figure}
\centering
\vspace{0.4cm}
\includegraphics[scale=0.4]{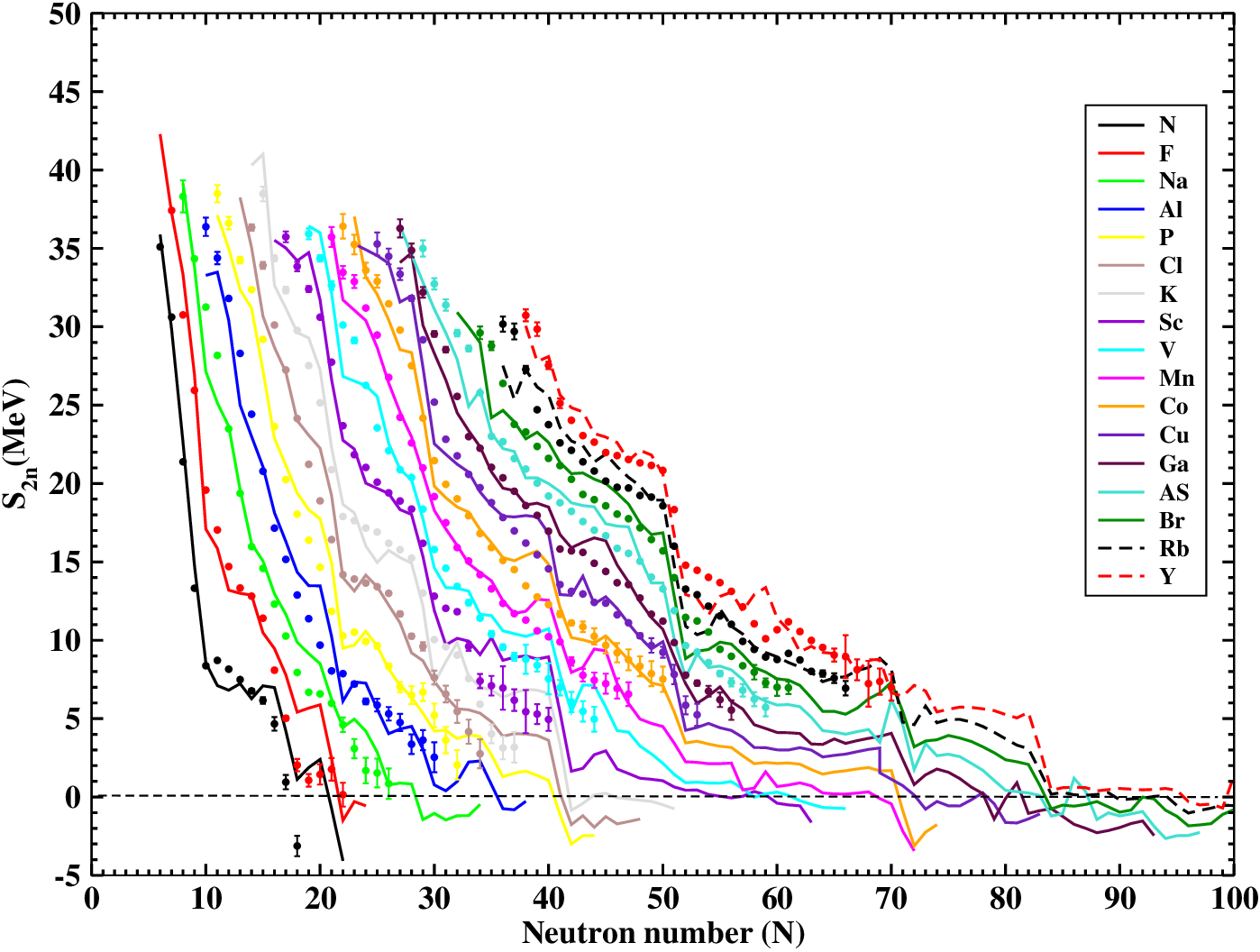}
\caption{(color online) Same as Fig.~\ref{sn1} but 
for odd-$Z$ isotopes from C to Zr.}
\label{sn3}
\end{figure}

\begin{figure}
\centering
\vspace{0.4cm}
\includegraphics[scale=0.4]{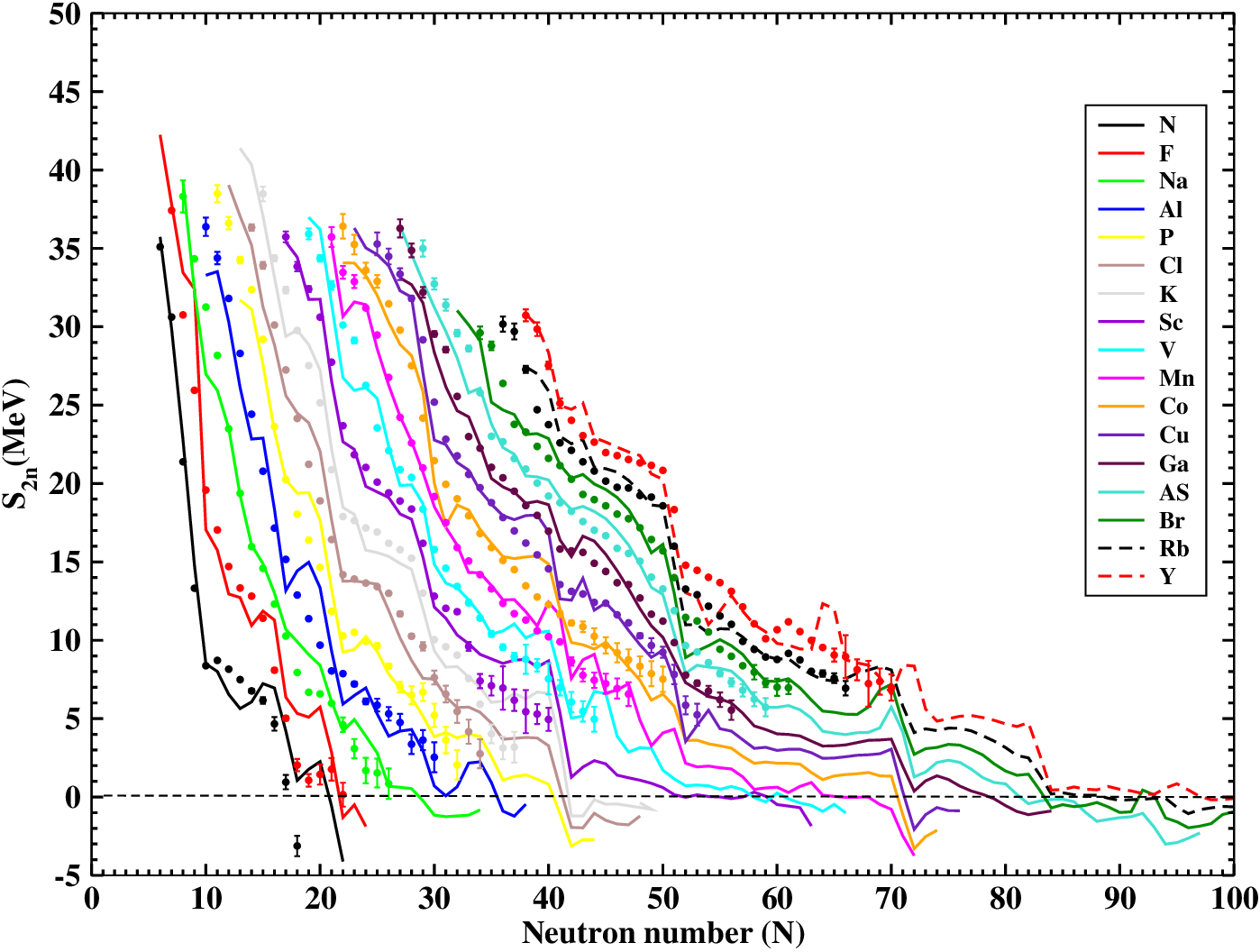}
\caption{(color online) Same as Fig.~\ref{sn3} but for NL3* parameterization.}
\label{sn4}
\end{figure}

\begin{table*}
\vspace{0.4cm}
\caption{The quantities nuclear asymmetry parameter ($\delta$=$N-Z/A$), BE/A(MeV), $S_{2n}$(MeV), $\beta_2$ and 
$\Delta r$(fm), calculated using  RMF theory with NL3 parameterization, are listed for two neutron drip line nuclei.}
\renewcommand{\tabcolsep}{0.14cm}
\renewcommand{\arraystretch}{1.1}
\small
\begin{tabular}{|cccccc|cccccc|}
\hline
Nuclei &$\delta$&BE/A&$S_{2n}$&$\beta_2$&$\Delta r$   &Nuclei&$\delta$&BE/A&$S_{2n}$&$\beta_2$&$\Delta r$ \\
\hline
$^{23}$C  &0.478&5.373&2.009  &0.062&0.934&$^{27}$N  &0.481&5.613&2.392 &0.005&0.884\\
$^{29}$O  &0.448&6.121&0.394  &0.043&0.887&$^{30}$F  &0.400&6.598&1.638 &0.045&0.744\\
$^{34}$Ne &0.459&5.928&0.760  &0.498&0.878&$^{39}$Na &0.436&6.260&0.833 &0.463&0.800\\
$^{41}$Mg &0.415&6.570&1.407  &0.449&0.807&$^{48}$Al &0.458&6.131&0.711 &0.104&0.943\\
$^{49}$Si &0.429&6.553&1.468  &0.015&0.883&$^{55}$P  &0.455&6.313&1.063 &0.006&0.867\\
$^{56}$S  &0.429&6.650&1.456  &0.055&0.801&$^{58}$Cl &0.414&6.825&0.790 &0.075&0.777\\
$^{59}$Ar &0.389&7.129&2.279  &0.039&0.715&$^{60}$K  &0.367&7.405&3.214 &0.003&0.668\\
$^{73}$Ca &0.444&6.522&0.110  &0.114&0.864&$^{76}$Sc &0.447&6.490&0.022 &0.127&0.918\\
$^{79}$Ti &0.443&6.550&0.134  &0.165&0.924&$^{80}$V  &0.425&6.774&0.493 &0.207&0.860\\
$^{87}$Cr &0.448&6.516&0.131  &0.159&0.959&$^{93}$Mn &0.462&6.364&0.323 &0.094&0.972\\
$^{95}$Fe &0.463&6.499&0.042  &0.064&0.950&$^{97}$Co &0.443&6.655&1.689 &0.001&0.945\\
$^{98}$Ni &0.433&6.847&3.031  &0.001&0.901&$^{101}$Cu&0.426&6.839&0.126 &0.077&0.881\\
$^{108}$Zn&0.444&6.622&0.177  &0.256&0.943&$^{110}$Ga&0.436&6.713&0.126 &0.249&0.906\\
$^{112}$Ge&0.446&6.808&0.436  &0.242&0.981&$^{115}$As&0.426&6.825&0.261 &0.235&0.862\\
$^{117}$Se&0.419&6.917&0.583  &0.007&0.764&$^{118}$Br&0.407&7.045&0.756 &0.010&0.738\\
$^{121}$Kr&0.404&7.053&0.218  &0.126&0.785&$^{126}$Rb&0.413&6.945&0.085 &0.175&0.865\\
$^{132}$Sr&0.424&6.794&0.187  &0.174&0.947&$^{134}$Y&0.418&6.852&0.013 &0.245&0.866\\
$^{135}$Zr&0.407&6.964&0.325  &0.259&0.820&          &     &     &      &     &     \\
\hline
\end{tabular}
\label{tab2}
\end{table*}

\begin{table*}
\vspace{0.4cm}
\caption{Same as Table~\ref{tab2}  but for NL3* parameterization.}
\renewcommand{\tabcolsep}{0.14cm}
\renewcommand{\arraystretch}{1.1}
\small
\begin{tabular}{|cccccc|cccccc|}
\hline
Nuclei &$\delta$&BE/A&$S_{2n}$&$\beta_2$&$\Delta r$   &Nuclei&$\delta$&BE/A&$S_{2n}$&$\beta_2$&$\Delta r$ \\
\hline
$^{23}$C  &0.478&5.306&1.954  &0.096&0.945&$^{27}$N  &0.481&5.561&2.255 &0.005&0.897\\
$^{29}$O  &0.448&6.067&0.499  &0.108&0.907&$^{30}$F  &0.400&6.488&2.844 &0.189&0.745\\
$^{34}$Ne &0.412&6.442&1.907  &0.462&0.734&$^{39}$Na &0.436&6.208&0.534 &0.463&0.815\\
$^{41}$Mg &0.415&6.514&1.616  &0.458&0.774&$^{48}$Al &0.458&6.078&0.918 &0.112&0.955\\
$^{49}$Si &0.429&6.499&1.221  &0.051&0.893&$^{55}$P  &0.455&6.265&0.815 &0.006&0.882\\
$^{56}$S  &0.429&6.606&1.198  &0.055&0.815&$^{58}$Cl &0.414&6.775&0.597 &0.099&0.797\\
$^{59}$Ar &0.390&7.088&2.137  &0.075&0.720&$^{60}$K  &0.367&7.361&3.381 &0.047&0.678\\
$^{71}$Ca &0.437&6.562&0.030  &0.115&0.903&$^{72}$Sc &0.417&6.797&0.149 &0.133&0.811\\
$^{79}$Ti &0.443&6.503&0.077  &0.158&0.942&$^{80}$V  &0.425&6.728&0.473 &0.195&0.880\\
$^{86}$Cr &0.442&6.550&0.158  &0.159&0.963&$^{90}$Mn &0.444&6.528&0.041 &0.094&0.964\\
$^{95}$Fe &0.453&6.454&0.275  &0.066&0.949&$^{97}$Co &0.443&6.614&1.329 &0.001&0.964\\
$^{98}$Ni &0.429&6.807&2.656  &0.002&0.919&$^{100}$Cu&0.420&6.873&0.351 &0.077&0.886\\
$^{101}$Zn&0.406&7.025&1.341  &0.256&0.837&$^{109}$Ga&0.431&6.732&0.169 &0.242&0.919\\
$^{112}$Ge&0.429&6.763&0.438  &0.241&0.894&$^{114}$As&0.421&6.841&0.164 &0.248&0.824\\
$^{117}$Se&0.419&6.863&0.177  &0.125&0.789&$^{118}$Br&0.407&6.997&0.437 &0.063&0.764\\
$^{121}$Kr&0.405&7.009&0.094  &0.126&0.805&$^{125}$Rb&0.408&6.962&0.110 &0.165&0.862\\
$^{131}$Sr&0.420&6.809&0.012  &0.174&0.952&$^{134}$ Y&0.426&6.724&0.035 &0.245&1.053\\
$^{135}$Zr&0.407&6.923&0.333  &0.259&0.968&          &     &     &      &     &     \\
\hline
\end{tabular}
\label{tab3}
\end{table*}

\begin{table*}
\centering
\vspace{0.4cm}
\caption{The possible number of neutron are listed for two-neutron drip line nuclei produced 
in this work by RMF model with NL3 and NL3* parameterizations. Comparison of present work has 
been made with prediction of spherical RCHB~\cite{xia2018}, available experimental data~\cite{Wang2017} and with macroscopic FRDM data~\cite{Moller}}.
\renewcommand{\tabcolsep}{0.15cm}
\renewcommand{\arraystretch}{1.1}
\small
\begin{tabular}{|cccccc|}
\hline
Nuclei &RMF+NL3&RMF+NL3*&Experiment&RCHB+PC-PK1&FRDM \\
\hline
C	&	17	&	17	&	16	&		&		\\
N	&	20	&	20	&	17	&		&		\\
O	&	21	&	21	&	17	&		&	19	\\
F	&	21	&	21	&		&	23	&	23	\\
Ne	&	24	&	24	&		&	32	&	24	\\
Na	&	28	&	28	&		&	35	&	27	\\
Mg	&	29	&	29	&		&	35	&	28	\\
Al	&	35	&	35	&		&	37	&	29	\\
Si	&	35	&	35	&		&	39	&	32	\\
P	&	40	&	40	&		&	41	&	35	\\
Se	&	40	&	40	&		&	41	&	36	\\
Cl	&	41	&	41	&		&	42	&	40	\\
Ar	&	41	&	41	&		&	45	&	43	\\
K	&	41	&	41	&		&	59	&	46	\\
Ca	&	53	&	51	&		&	61	&	49	\\
Sc	&	55	&	51	&		&	62	&	50	\\
Ti	&	57	&	57	&		&	62	&	51	\\
V	&	57	&	57	&		&	65	&	52	\\
Cr	&	65	&	64	&		&	69	&	54	\\
Mn	&	68	&	65	&		&		&	54	\\
Fe	&	69	&	69	&		&	71	&	57	\\
Co	&	70	&	70	&		&	71	&	59	\\
Ni	&	70	&	70	&		&	71	&	64	\\
Cu	&	72	&	71	&		&	71	&	67	\\
Zn	&	78	&	71	&		&	73	&	70	\\
Ga	&	78	&	78	&		&	79	&	72	\\
Ge	&	80	&	80	&		&	83	&	75	\\
As	&	81	&	81	&		&	85	&	82	\\
Se	&	83	&	83	&		&	94	&	82	\\
Br	&	83	&	83	&		&	96	&	82	\\
Kr	&	85	&	85	&		&	100	&	83	\\
Rb	&	89	&	88	&		&	102	&	83	\\
Sr	&	94	&	93	&		&	111	&	83	\\
Y	&	95	&	95	&		&		&	83	\\
Zr	&	95	&	95	&		&		&	85	\\

\hline
\end{tabular}
\label{tab4}
\end{table*}

\begin{figure}
\centering
\vspace{0.4cm}
\includegraphics[scale=0.4]{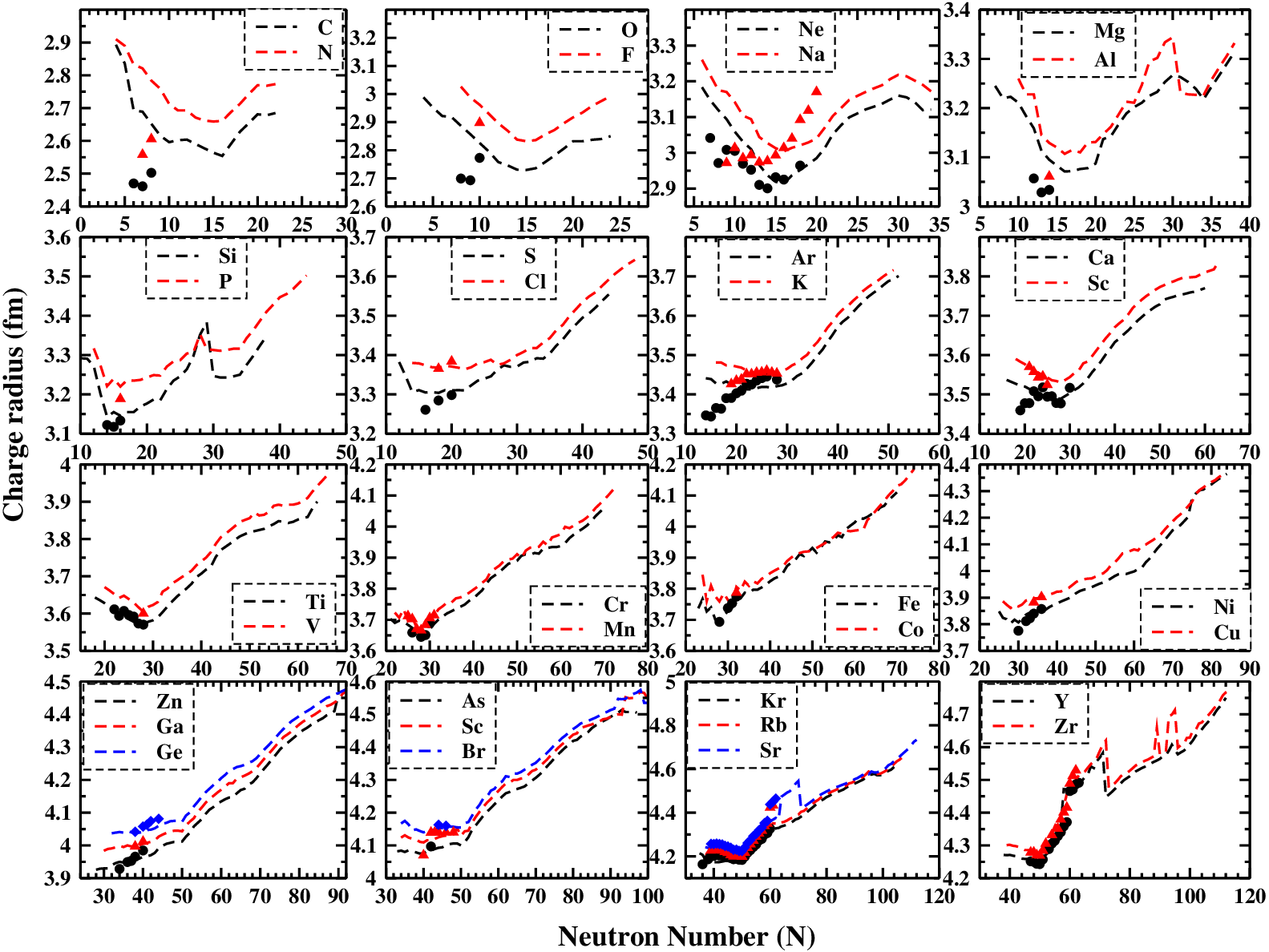}
\caption{(color online) The charge radius ($r_{c}$), calculated using  RMF theory with NL3 parameterization,
is plotted as a function of neutron number from C to Zr isotopes. 
Experimental data are represented by filled circles, triangles and diamonds.}
\label{rch1}
\end{figure}

\begin{figure}
\centering
\vspace{0.4cm}
\includegraphics[scale=0.4]{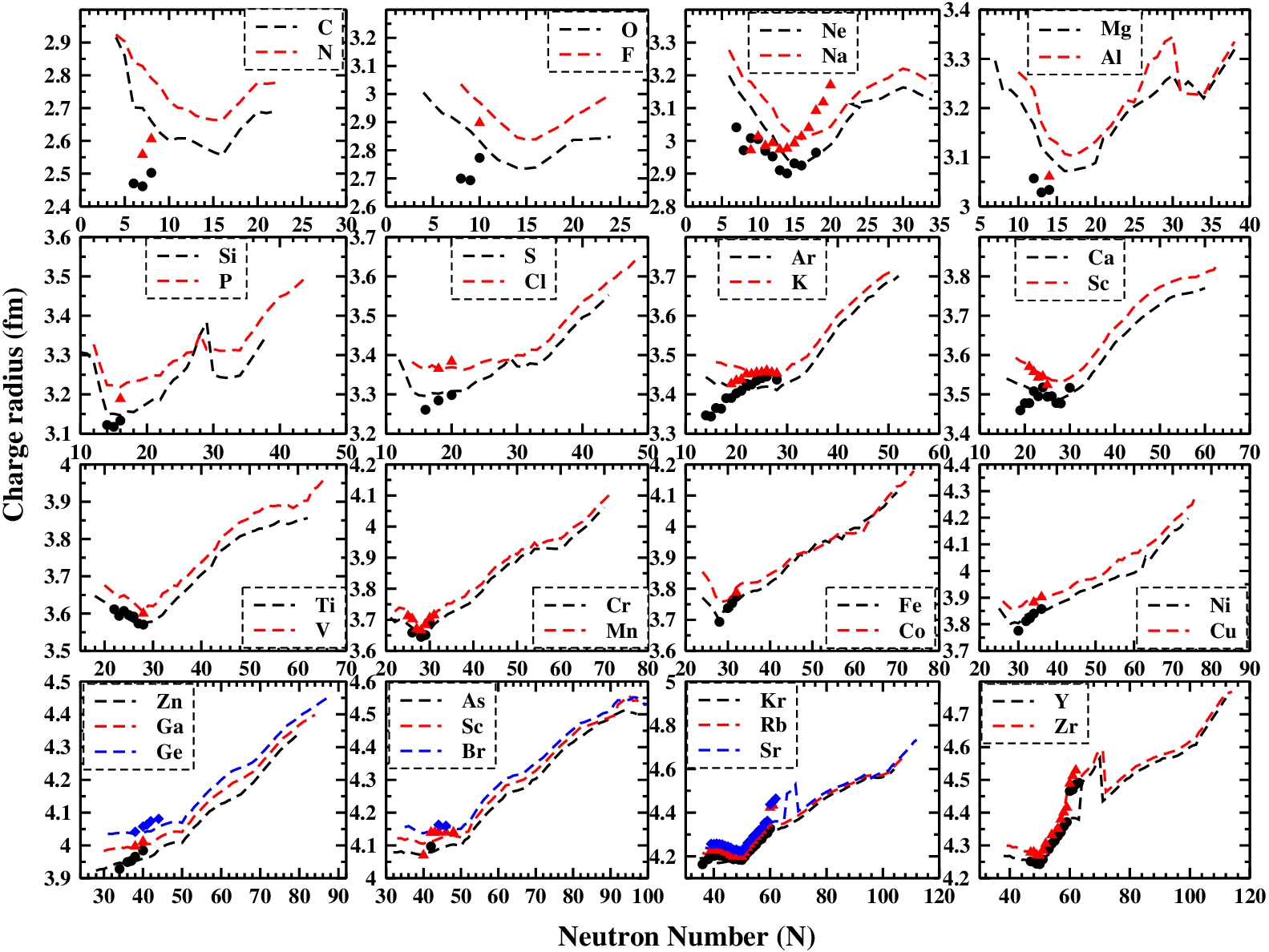}
\caption{(color online) Same as Fig.~\ref{rch1} but for NL3* parameterization.}
\label{rch2}
\end{figure}
\begin{figure}
\centering
\vspace{0.4cm}
\includegraphics[scale=0.4]{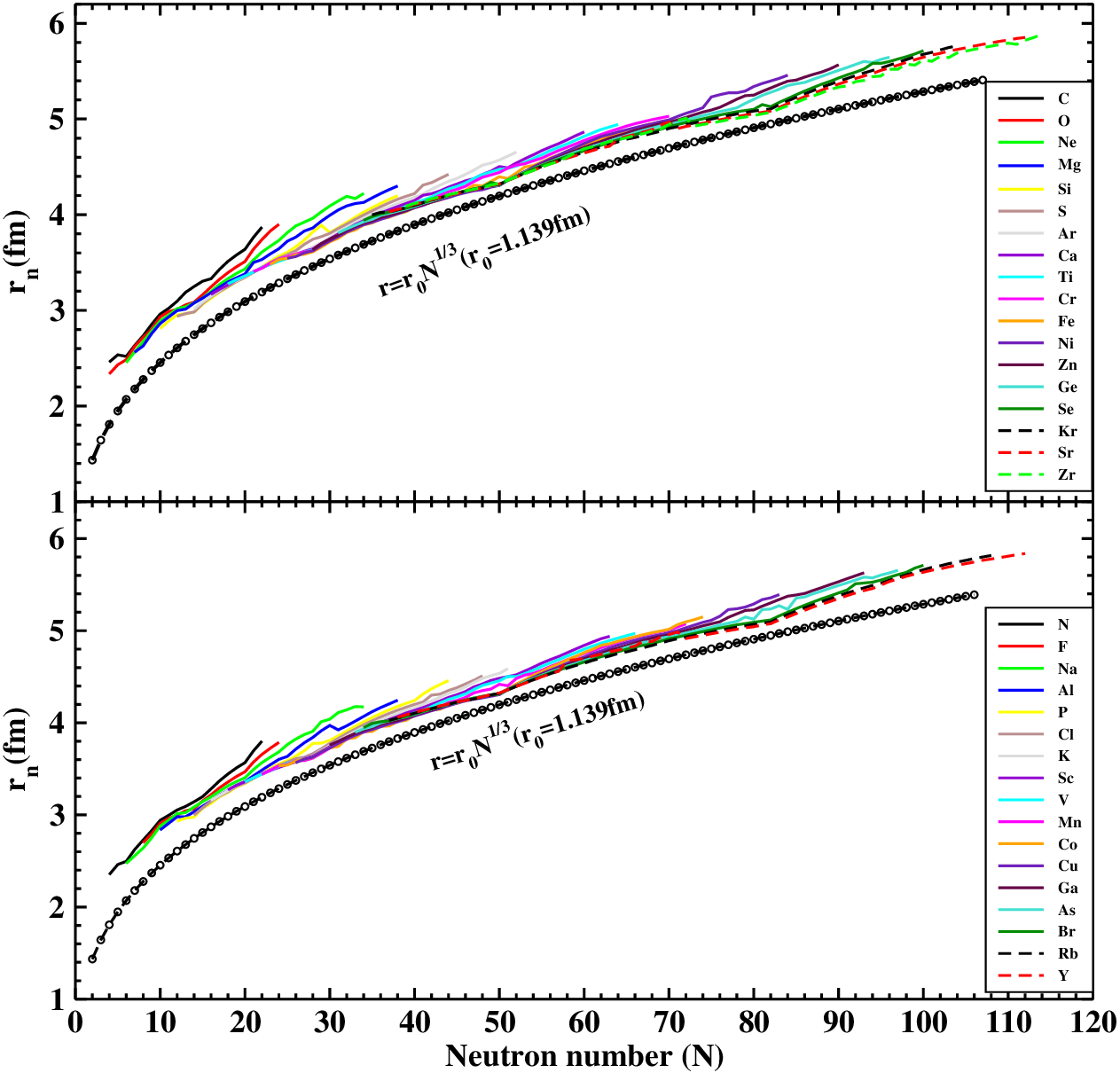}
\caption{(color online)The root mean square neutron radii, 
calculated using  RMF theory with NL3 parameterization,
is plotted as a function of neutron 
number for even-$Z$ isotopes (upper panel) and for odd-$Z$ isotopes (lower panel) from C to Zr.}
\label{rms1}
\end{figure}

\begin{figure}
\centering
\vspace{0.4cm}
\includegraphics[scale=0.4]{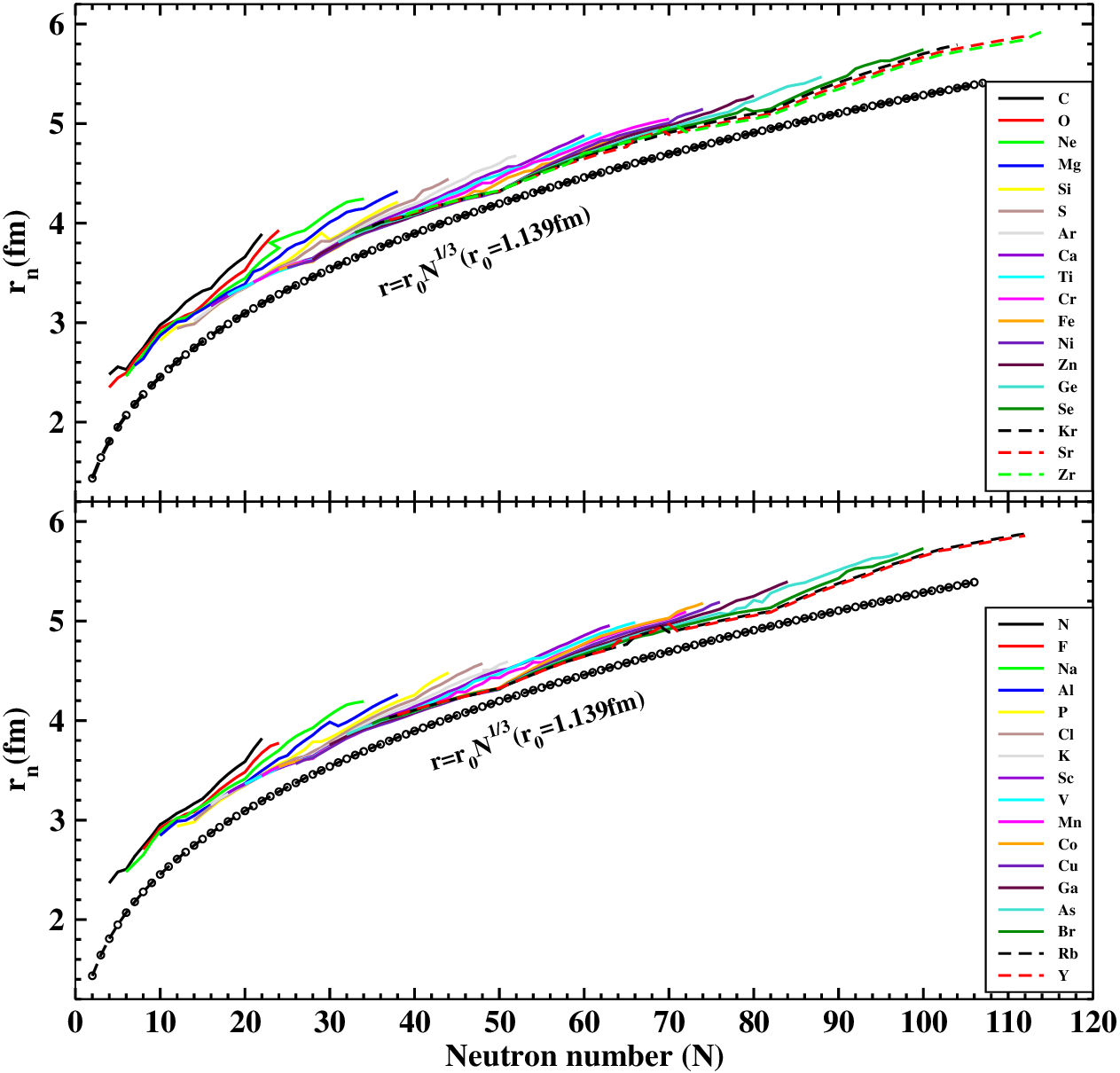}
\caption{(color online) Same as Fig.~\ref{rms1} but for NL3* parameterization.}
\label{rms2}
\end{figure}
\begin{figure}
\centering
\vspace{0.4cm}
\includegraphics[scale=0.4]{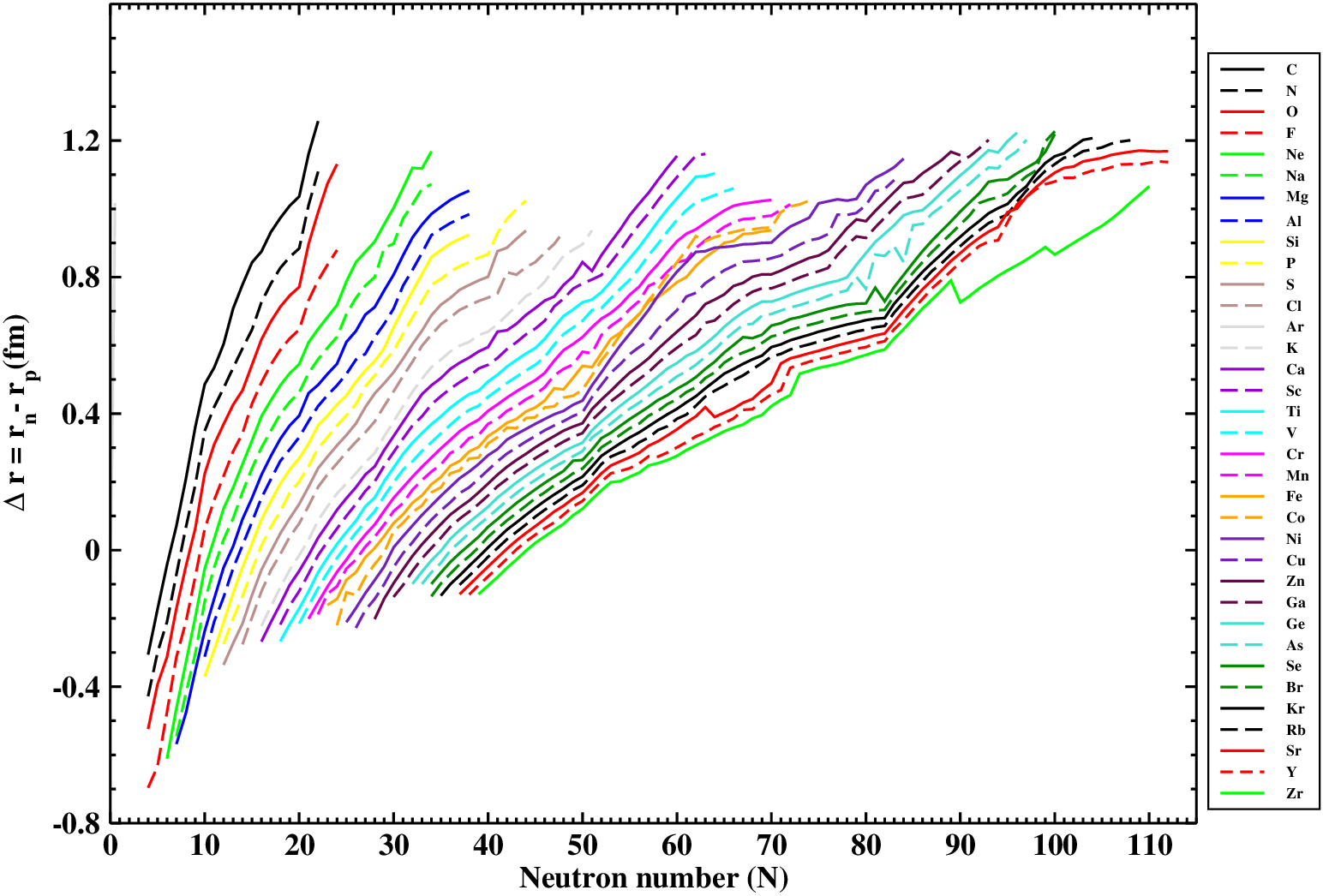}
\caption{(color online) Neutron skin-thickness from $\beta$ stable to drip line region, 
calculated using  RMF theory with NL3 parameterization, is given as a 
function of neutron number for nuclei from C to Zr. Even-$Z$ isotopes are represented by solid line whereas 
dashed line is used for showing the odd-$Z$ isotopes.}
\label{skin1}
\end{figure}

\begin{figure}
\centering
\vspace{0.4cm}
\includegraphics[scale=0.4]{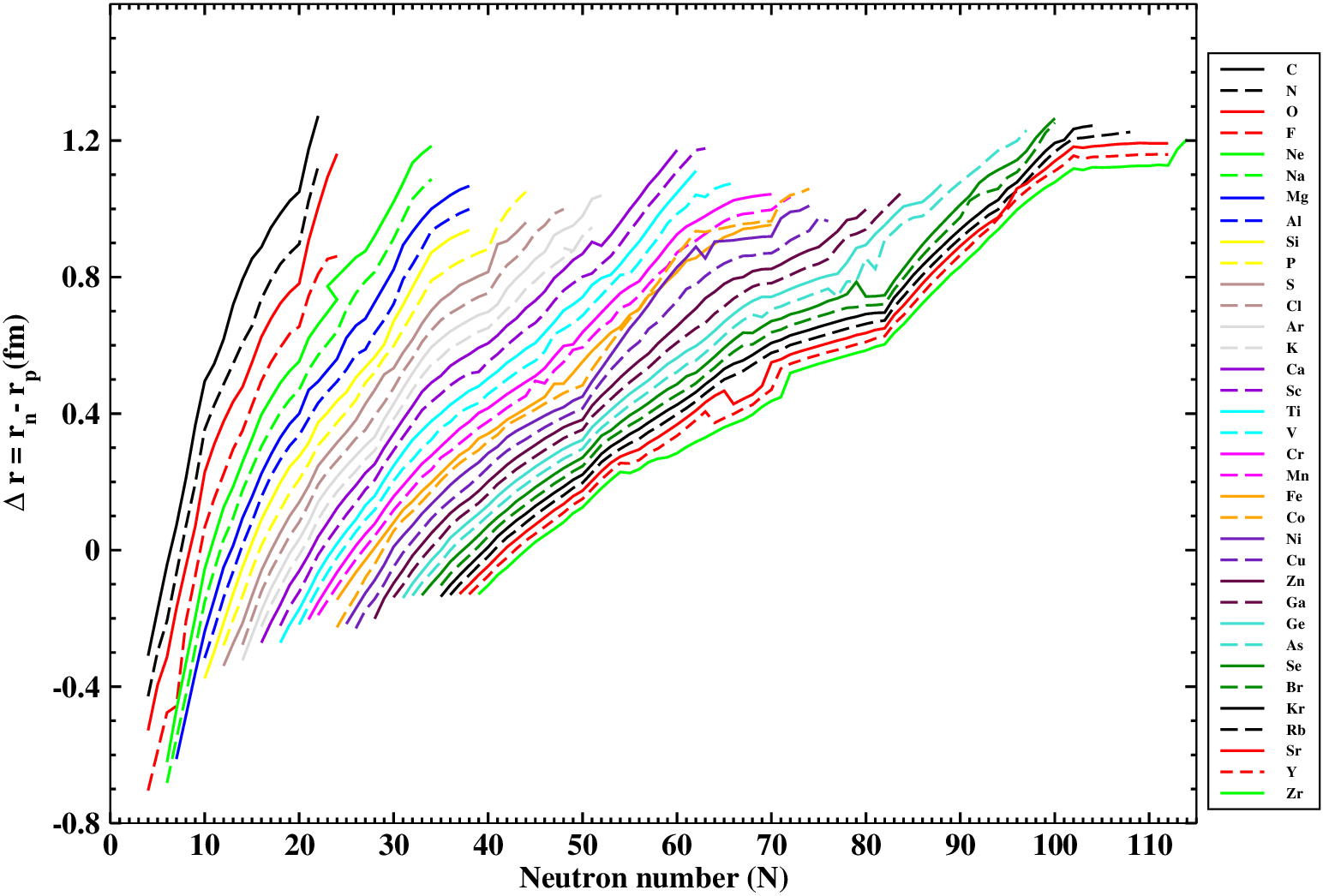}
\caption{(color online) Same as Fig.~\ref{skin1} but for NL3* parameterization.}
\label{skin2}
\end{figure}
\begin{figure}
\centering
\vspace{0.4cm}
\includegraphics[scale=0.4]{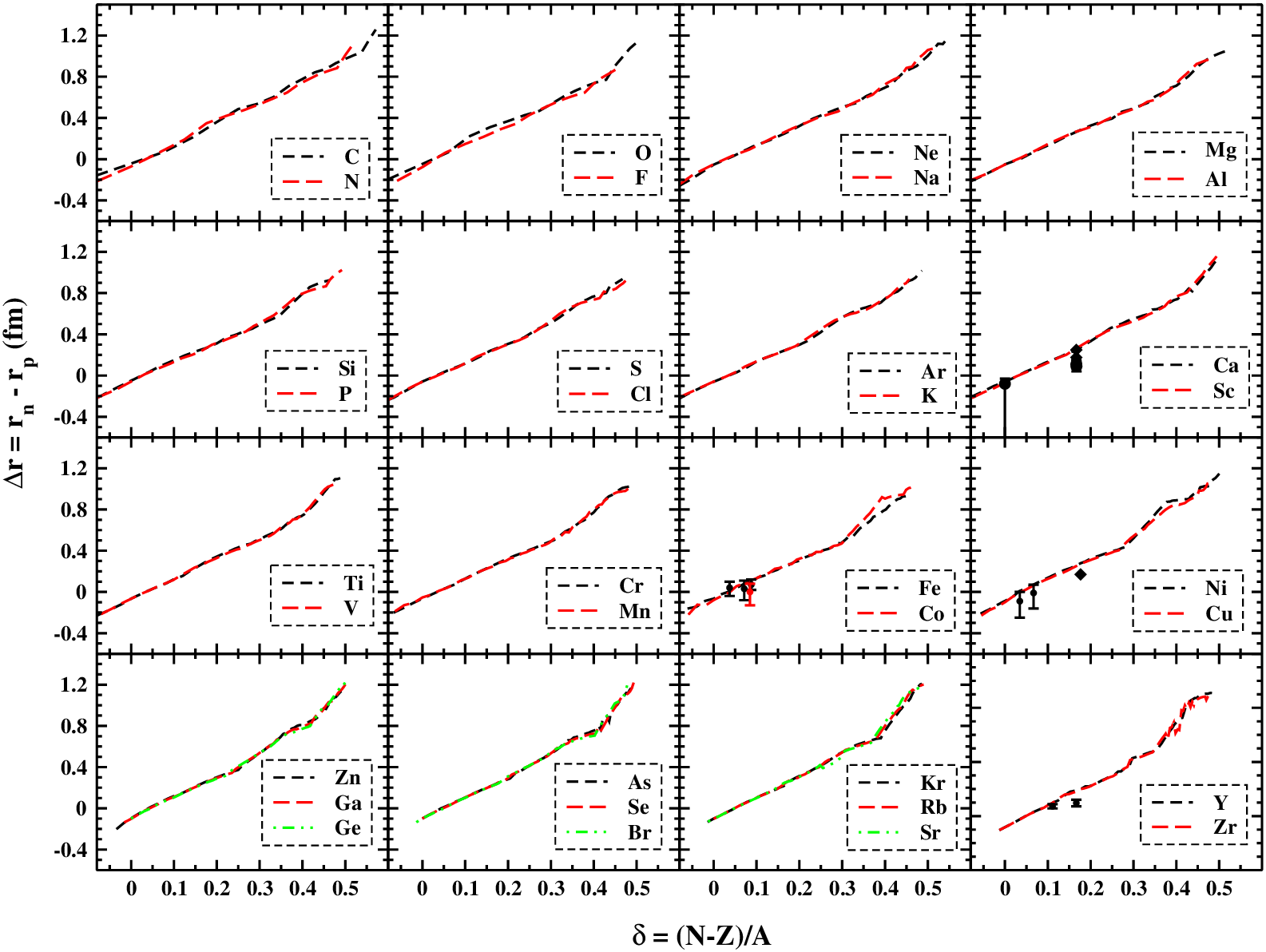}
\caption{(color online) Neutron skin-thickness from $\beta$ stable to drip line region,
calculated using  RMF theory with NL3 parameterization, is plotted as a 
function of asymmetry parameter ($\delta$ = $N-Z/A$) for the nuclei from C to Zr. 
Available experimental data are represented by filled circles with error bars.
Available earlier theoretical predictions are symbolized by diamonds or triangles.}
\label{iskin1}
\end{figure}

\begin{figure}
\centering
\vspace{0.4cm}
\includegraphics[scale=0.4]{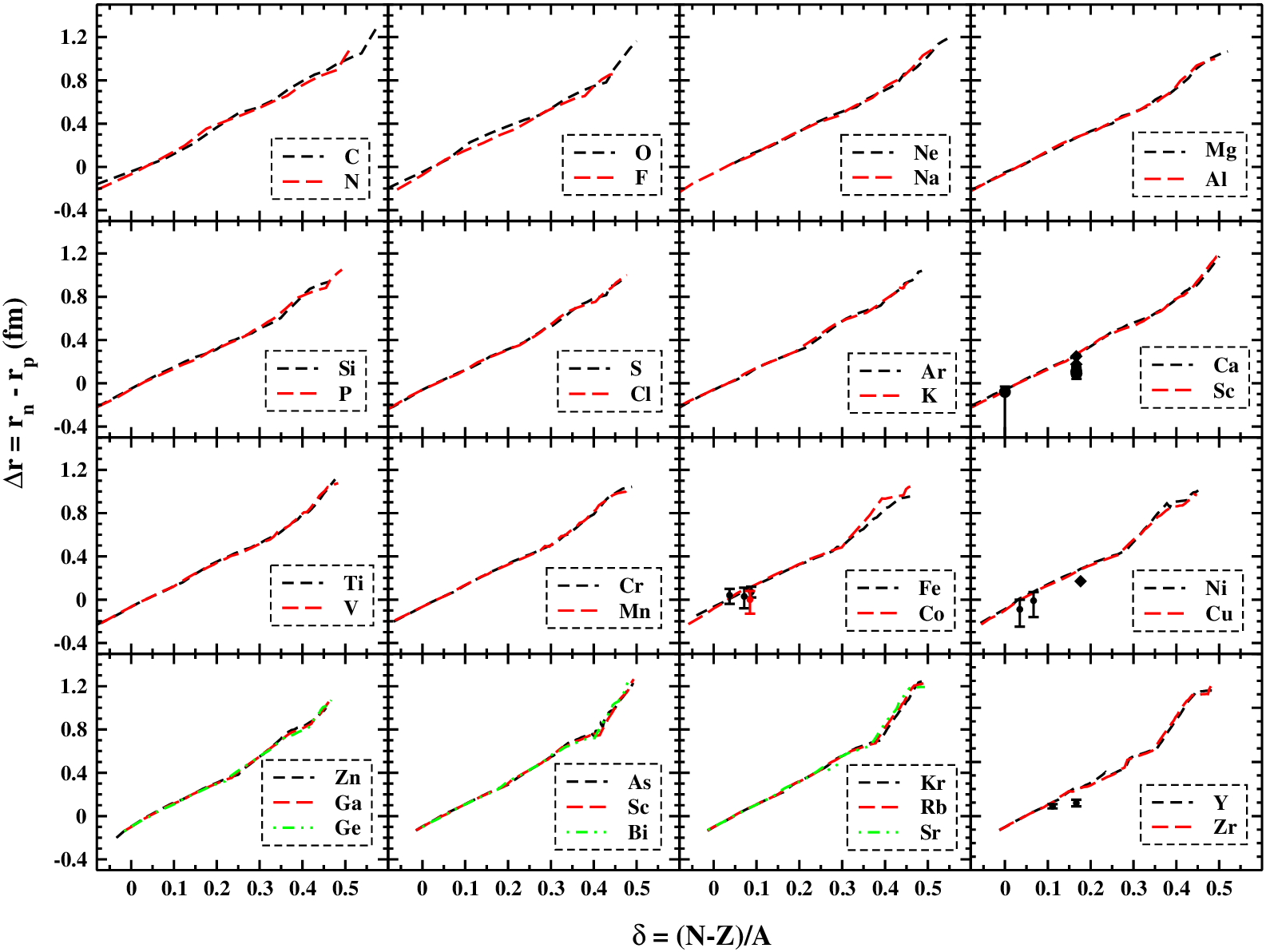}
\caption{(color online) Same as Fig.~\ref{iskin1} but for NL3* parameterization.}
\label{iskin2}
\end{figure}

\begin{figure}
\centering
\vspace{0.4cm}
\includegraphics[scale=0.45]{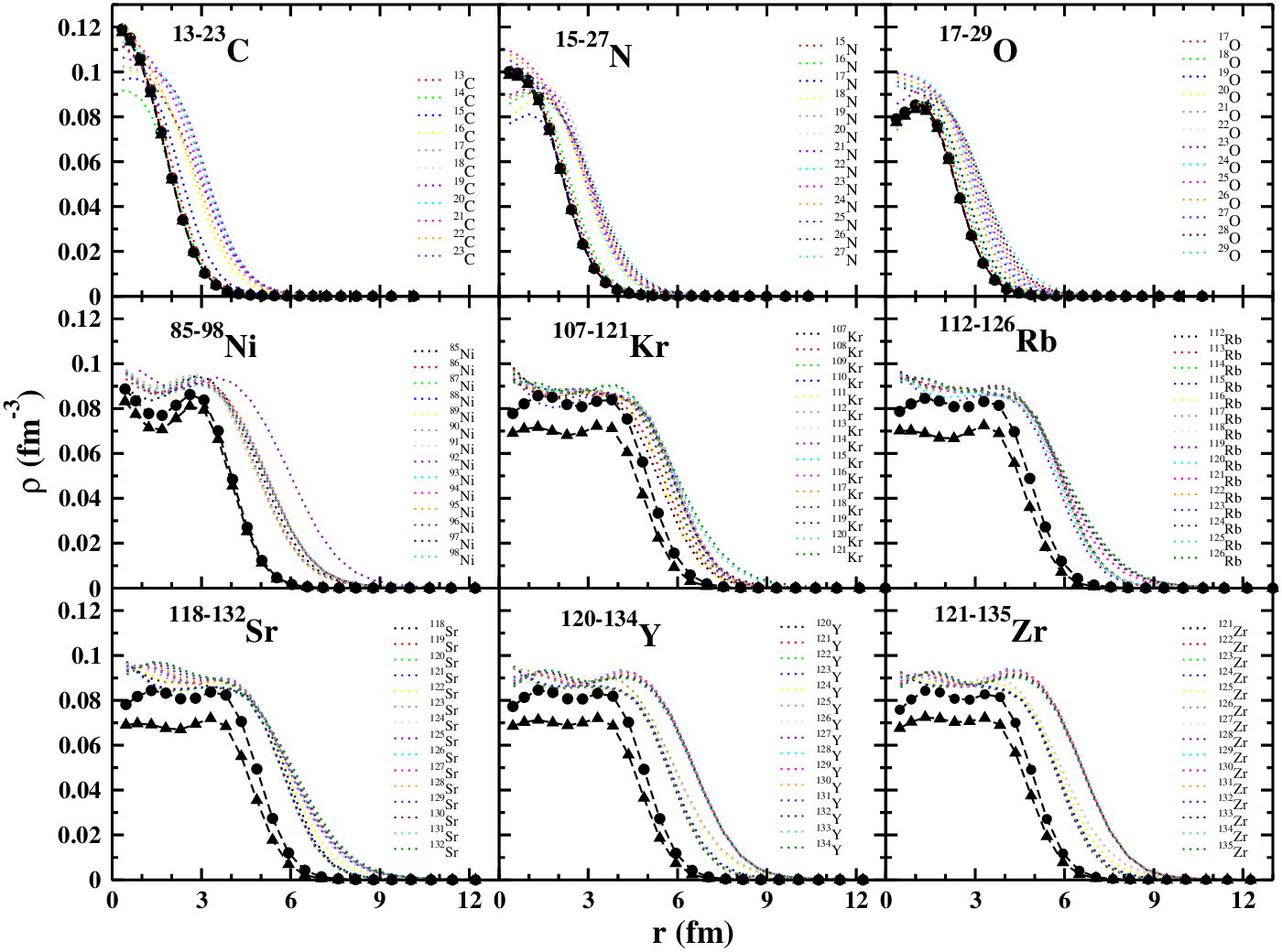}
\caption{(color online) Neutron density distributions for rich skin candidates 
(C, N, O, Ni, Kr, Rb, Sr, Y and Zr isotopic chain) as a function of radial parameter ($r$) is given.
Neutron (line with filled circles) and proton densities (line with filled triangles) for stable cases are also plotted.
These densities are calculated using  RMF theory with NL3 parameterization.}
\label{den1}
\end{figure}

\begin{figure}
\centering
\vspace{0.4cm}
\includegraphics[scale=0.4]{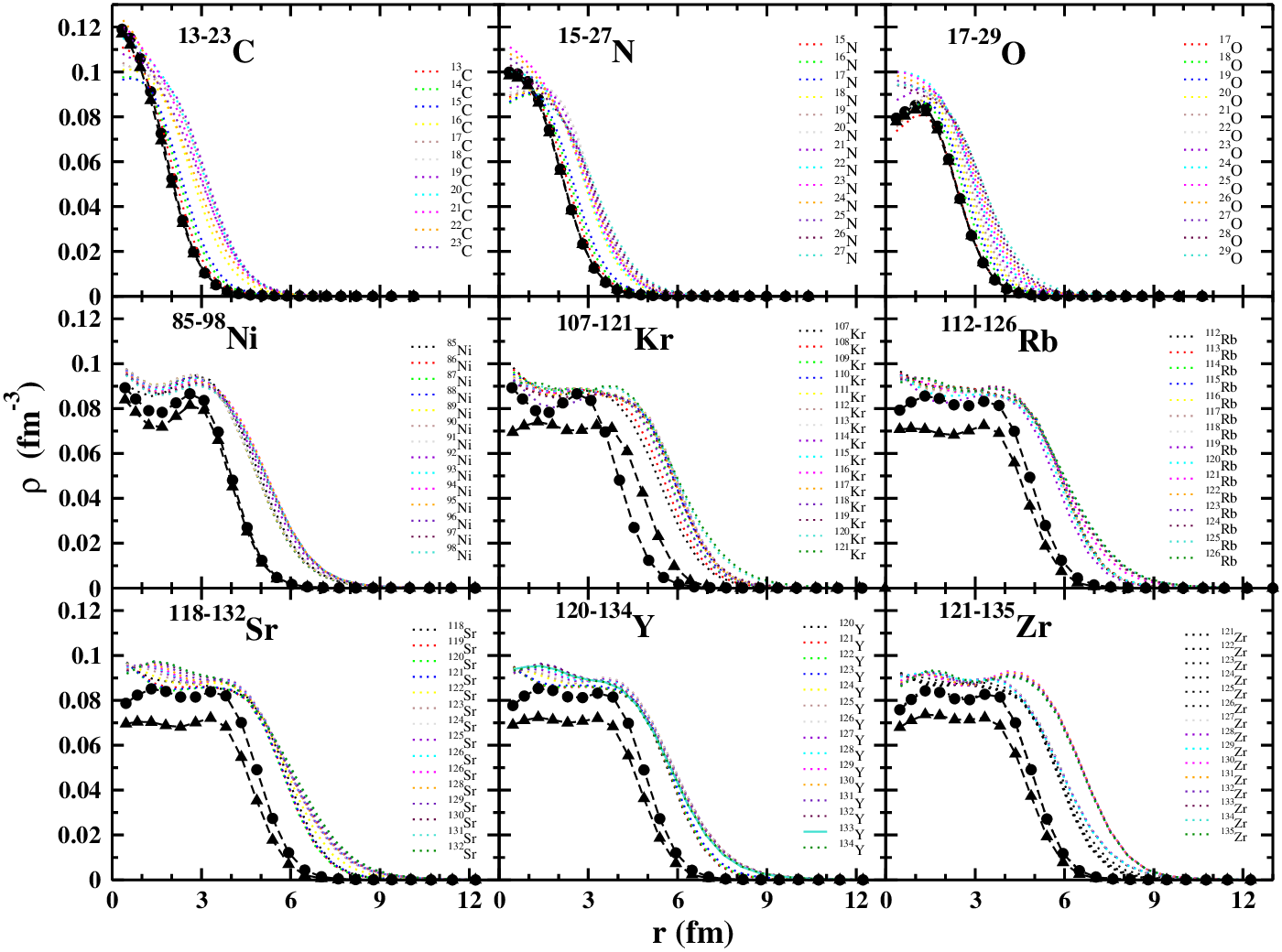}
\caption{(color online) Same as Fig.~\ref{den1} but for NL3* parameterization.}
\label{den2}
\end{figure}
\begin{figure}
\centering
\vspace{0.4cm}
\includegraphics[scale=0.45]{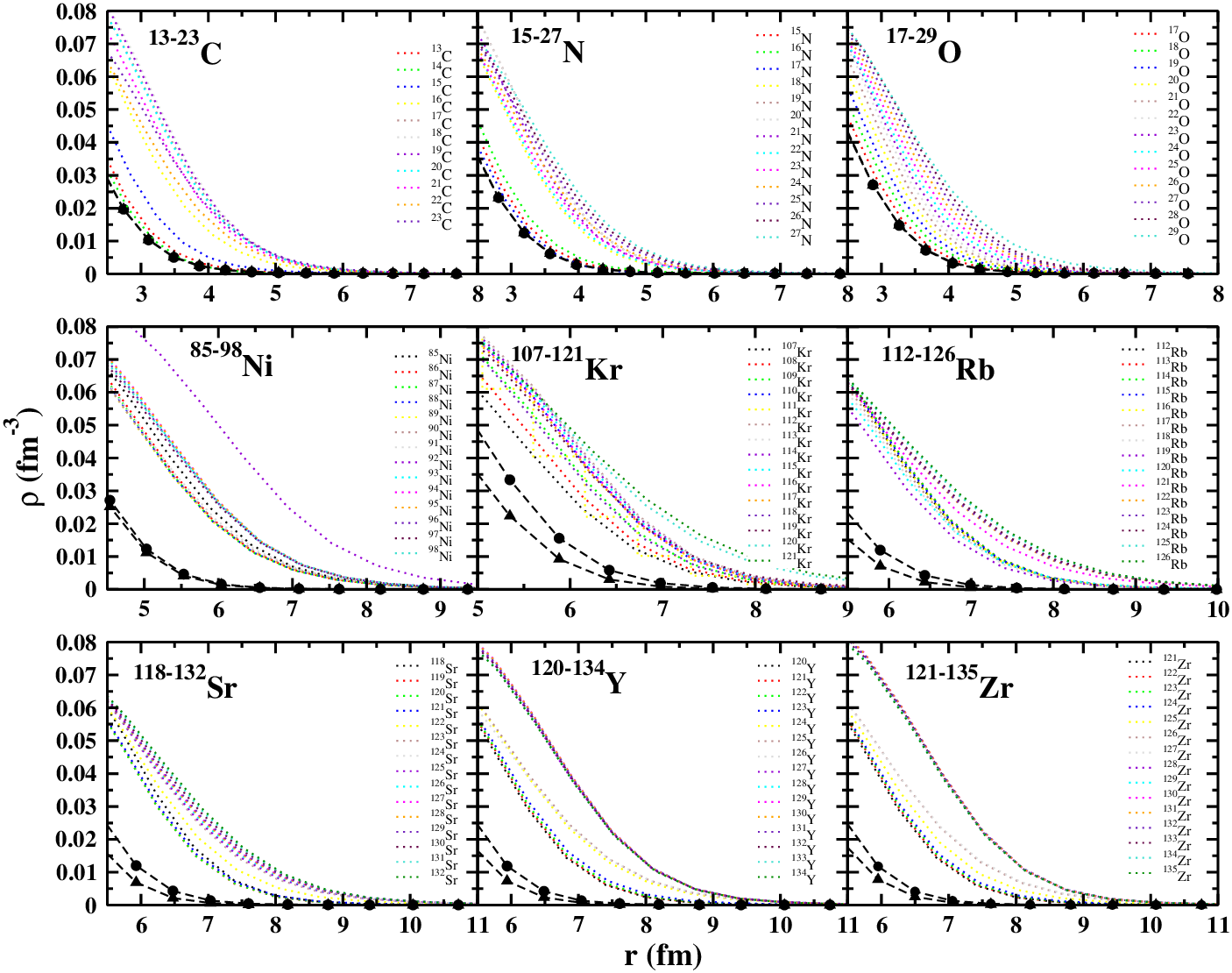}
\caption{(color online) Neutron density distributions for said rich 
skin candidates but for peripheral region 
in view to guide the long tails. These densities are calculated using  RMF theory with NL3 parameterization.}
\label{den3}
\end{figure}

\begin{figure}
\centering
\vspace{0.4cm}
\includegraphics[scale=0.4]{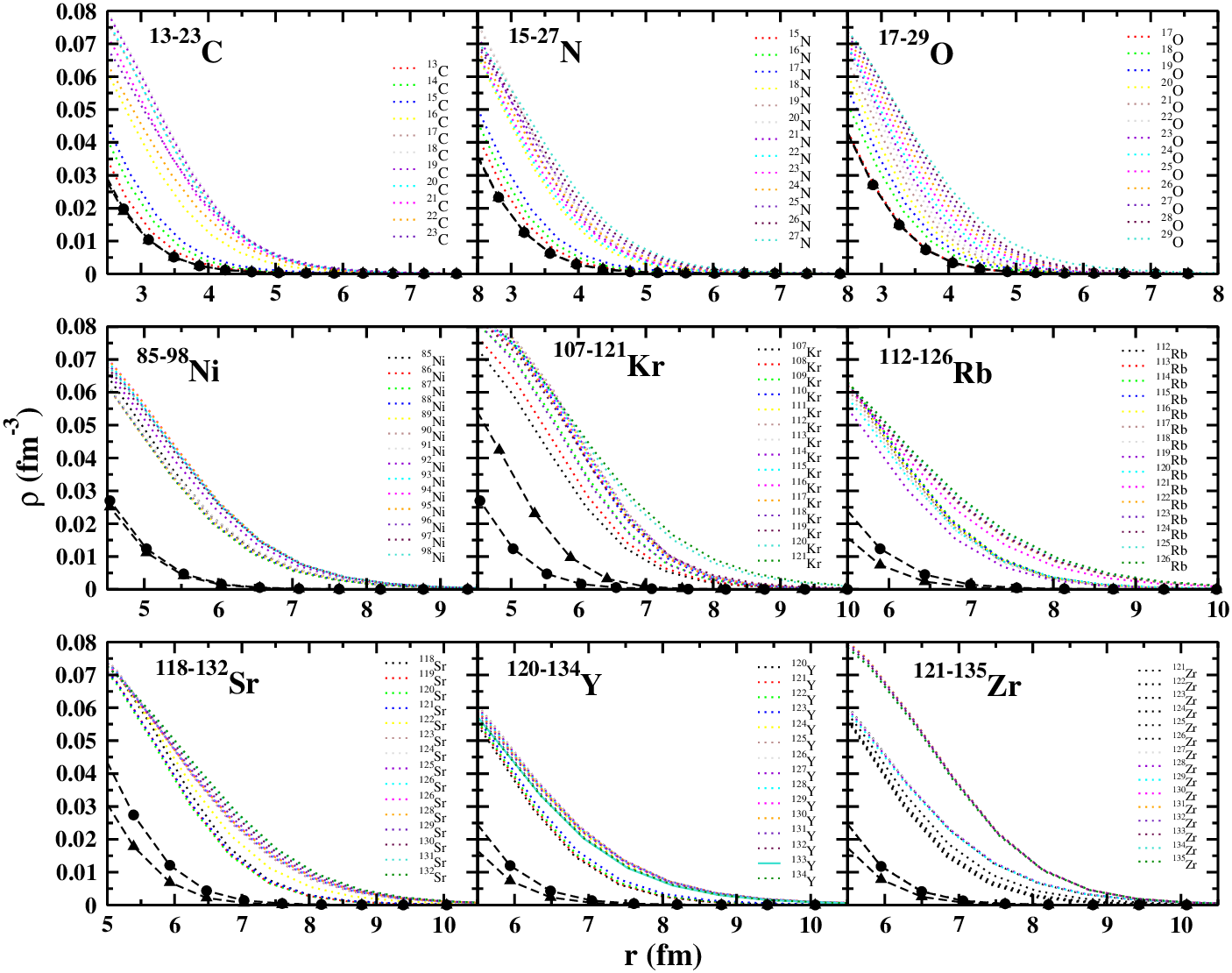}
\caption{(color online) Same as Fig.~\ref{den3}, but for NL3* parameterization.}
\label{den4}
\end{figure}

\subsection{Binding energy and separation energy}
The calculated binding energy in terms of binding energy 
per particle ($BE/A$) and two-neutron separation 
energies of the isotopic chain from carbon to zirconium isotopes are plotted 
in Figs.~\ref{be1}-\ref{sn4} with NL3 and NL3* parameterizations.
The highest value of $BE/A$ for the nuclide corresponds 
to the maximum stability and is 
known as the most stable nucleus. 
Further, with an increase in the number of neutrons with a 
fixed value of $Z$ so-called $N/Z$ ratio increases, 
and corresponding $BE/A$ monotonically decreases. 
This behaviour of the nuclides can be seen here 
in Fig.~\ref{be1},\ref{be2} and it must be generalized over the whole periodic table. 
Binding energy per particle obtained from the model and 
the available data from the experiment~\cite{Wang2017} is quite agreeable. 
However, the experimental data are not available up to the drip line region.
Parabolic shapes are produced for every isotopic chain and the most 
bound isotopes lie on the deepest of the curve.
Moreover, the two-neutron separation energies are estimated 
from the binding energies to explain the microscopic 
behaviour of the nuclei and other nuclear phenomena such as drip line, halo, or skin structure. 
The two-neutron separation energies are estimated using the following relation
\begin{equation}
 S_{2n}(N,Z) = BE(N,Z)-BE(N-2,Z) . 
\end{equation}
The halo nucleus has the minimum value of $S_{2n}$ on 
contrast a large value of neutron radius. 
Here, we have plotted the two-neutron separation energies 
for all considered isotopic series as a function of 
neutron number as given in figures~\ref{sn1} to ~\ref{sn4}. 
On the basis of $S_{2n}$, the drip line nucleus is anticipated  over the entire chain of isotopes.
For instance, the relativistic continuum Hartree-Bogolivbov (RCHB) theory reports the drip line 
nucleus to be $^{30}$O~\cite{meng2}, but our RMF with NL3 and NL3* marks $^{29}$O
as a drip nucleus, whereas the experimentally observed nucleus is $^{26}$O.
The drip line isotope for C is predicted as $^{24}$C within RCHB framework~\cite{poschl}, 
where our results suggest as $^{23}$C.
Also, the last bound isotope for Ca is predicted as $^{72}$Ca within RCHB~\cite{meng2}, 
HFB~\cite{fayans}, SHF~\cite{soojae,sagawa} methods 
and our result predicts $^{73}$Ca and 
$^{71}$Ca for NL3 and NL3* parametrizations, respectively. 
In the case of Ni,  $^{100}$Ni is predicted as a drip nucleus 
with RCHB~\cite{meng2} while present 
calculations suggest as $^{98}$Ni is the drip line nucleus with neutron number $N=$ 70 within both parameterizations.
This $N=$ 70 has appeared here as a semimagic number. 
Moreover, both the parameterizations produce consistent results with each other.
The disparities, in some cases, between RMF and RCHB are arises due to exclusion of 
continuum states in BCS pairing approximation within RMF(NL3/NL3*) framework. 

Here, in Table ~\ref{tab4}, we have enlisted the possible 
neutron number for the two-neutron drip line 
nuclei in this work and a comparison is established
between our work and the predictions 
of spherical RCHB~\cite{xia2018} with  PC-PK1 parameterization and macro-microscopic 
finite range droplet model (FRDM)~\cite{Moller}. The two-neutron drip lines for  
C, N, and O have also been determined using experimental two-neutron separation
energies and also listed in one of the columns of Table ~\ref{tab4}. 
The two neutron drip line found in this work
lies nearby to the available experimental evaluation but differs largely from the prediction 
of FRDM and spherical RCHB~\cite{xia2018}. 
Therefore, reliable prediction of drip line is a 
matter of discussion in nuclear physics because 
its predictions are not only model but interaction dependent also.

In the context of radii, the charge radius produced by RMF(NL3) 
or RMF(NL3*) matches 
with RCHB predictions~\cite{xia2018}.
For example, for $^{28}$O, the charge value is $r_c=$ 2.903 fm 
within RCHB whereas 2.832 fm and 2.837 fm for RMF(NL3) and RMF(NL3*).
Also, for $^{30}$F, $r_c=$ 2.977 fm within RCHB whereas 2.934 fm 
and 2.937 fm for present NL3 and NL3* interactions.
For $^{49}$Si nucleus, the value of $r_c$ within RCHB is 3.332 fm 
on the other hand charge radius comes out 
to be 3.275 fm and 3.272 fm in present RMF calculations.
In case of $^{95}$Fe,  the continuum method produces 
the charge radius around $r_c=$ 4.149 fm, whereas 
BCS mean-field predicts as 4.097 fm and 4.096 fm. 
In the same way, for  $^{135}$Zr the amount of 
charge radius $r_c=$ 4.643 fm for continuum 
method and 4.592 fm and 4.589 fm for mean-field 
method in present calculations.
Therefore, it is evident that the continuum method 
like RCHB produces the upper limit of charge radius 
than RMF (NL3/NL3*) predictions.

It is worth mentioning that a nucleus may have any possible configuration from 
three kinds; oblate or spherical or prolate.
Nuclei optimize the energy corresponding with their shape configuration.
Sometime nucleus exists in two or more possible shape 
configurations with almost the same energies which is 
termed as  shape coexistence in literature.
Here, the shape of the nuclei for considered 
isotopic series comes out to be spherical or 
prolate by examining the deformation parameter. 
There is no such oblate shaped. Even some of 
the drip line nuclei have large 
prolate shaped and these results are 
consistent with experimental values~\cite{Raman}.
For example, $^{34}$Ne, $^{39}$Na, and $^{41}$Mg 
drip line nuclei have a large value of 
$\beta_2 \approx$ 0.4 within both parameterization 
and therefore predict a large deformed shape.
Other than large or mild prolate shaped, 
drip line nuclei have a spherical configuration.  
Moreover, both the interactions produce 
consistent results with each other.


Separation energy is supposed to be the first signature 
in distinguishing the magic numbers for an isotopic series. 
The extended shell gaps in the single-particle energy levels 
are identified as magic numbers in nuclei. 
This suggests an abrupt fall in neutron separation energy 
indicates the signature of the neutron magic number. 
In general, the separation energy decreases smoothly 
with increasing neutron numbers.
But, there are found to be kinks at $N=$ 20, 40, 50, 70, 
and 82 in figures~\ref{sn1},\ref{sn2}, \ref{sn3} and~\ref{sn4}, which 
are all traditional magic or semimagic numbers. 
However, neutron magic number $N=$ 28 is no longer 
seen and disappear in all isotopes.
The in held spin-orbit splitting in 
RMF formalism gives the strength to this model 
to reproduce the experimental magic number.

\subsection{Charge, neutron radii and neutron skin-thickness}
The charge radius and the way neutrons are distributed in the 
nucleus are the fundamental properties to find 
the dimension of the nucleus.
Experimentally, charge radii for stable nuclei are obtained through electron 
scattering with a high level of accuracy. 
Such studies provide information about a few of the foremost principal  
properties of nuclei, such as size, surface thickness, 
shell structure, and distribution of nucleons.
Contrary to the nuclear charge densities, we still have 
a circumscribed knowledge about neutron densities. 
And therefore, in nuclear physics, determining $r_n$ of 
a nucleus is a problem of fundamental significance.
In order to illustrate the size and probably halo/skin 
structure of the considered isotopic series, 
rms charge, rms neutron radii, and neutron skin thickness are 
calculated within the RMF theory from $\beta$ 
stable to the neutron drip line 
region and plotted in figures~\ref{rch1}-\ref{rms2}.
It is evident from Figs~\ref{rch1},\ref{rch2} that 
the $r_{c}$ is smaller for the stable 
isotopes in comparison to the isotopes lie near 
proton or neutron drip line regions.
It is the point in the curve, where asymmetry and 
Coulomb terms are in extremely balancing 
positions, giving an extremely stable isotope. 
The charge radius is seen as least for $^{28}$Si 
with $N=$ 14, owing to the shell effect and 
formed the most stable nucleus among Si isotopic chain.
Before and after $N=$ 14 the charge radius increases 
with changing in neutron number.
This trend of $r_{c}$ is followed by all considered 
isotopes as manifested in Figs.~\ref{rch1},\ref{rch2}.
The abrupt increase in $r_{c}$ leads to changes 
in the shape of the isotopes than previous ones.
Calculated theoretical values of charge radius for these 
isotopes agree well with available experimental data~\cite{Angeli}. 
However, a deviation for light nuclei 
is observed but an excellent consistency  
between our theoretical results and experimental 
data is noticed for the medium mass range of nuclei.

In Figs.~\ref{rms1},\ref{rms2} the determined rms 
neutron radii are plotted against the number of neutrons.
It is fascinating to see that $r_n$ follows 
the stability curve nicely for $\beta$ stable 
region and makes the deviation for exotic drip line region.
The stability curve is computed by the empirical formula $r=r_0N^{1/3}$.  
The sharp increase in the slope of $r_n$ at the drip line region indicates 
the larger neutron radius than the normal 
trend and probably skin structure shall appear.
A deviation of $r_n$ from empirical prediction is seen for the isotopes 
C, N, O, Ni, Kr, Sr, Rb, Y, and Zr  in the exotic mass region.
Of course, these rises in $r_n$ are due to valence 
neutrons adjusted to nuclei and forming the skin/halo structure.
This fact strongly supports the possibility for 
the existence of the neutron skin/halo 
in exotic C, N, O, Ni, Kr, Sr, Rb, Y, and Zr isotopes.
The outcomes of these results are in favour of earlier predictions~\cite{meng2,meng1,meng3,meng4,meng5}. 
However, deviation of $r_n$ for C, N, O are under 
suspicion because RMF model doesn't appear well in case of  
light mass nuclei due to its mean-field nature~\cite{patra1993}.

In figures~\ref{skin1},~\ref{skin2}, we show the neutron skin thickness as 
a function of neutron number $N$ for the considered isotopic chains. 
The magnitude of skin thickness increasing systematically with the number of neutrons 
within the isotopes. The gradual increment in the neutron skin 
may be described as the redistribution of the nucleons 
with  the additament of neutrons to a stable nucleus. 
The slope of the thickness is larger for some isotopes, 
for example, C, N, O, Ni, Kr, Rb, Sr, Y, and Zr 
due to their halo nature as predicted earlier by authors~\cite{meng2,meng1,meng3,meng4,meng5}.
It is to be noted that Ni with $N$ $>$ 50 
and Kr, Rb, Sr, Y, Zr isotopes with $N$ $>$ 82 show a 
sharp increase in skin thickness and a large 
decrease in two neutron separation energies. 
For example, Ni with $N=$ 50 has the value of 
$S_{2n}$= 9.468 MeV while this amount goes to 3.031 MeV at $N$= 52.
Neutron skin thickness varies in magnitude 
from stable to neutron drip line region. 
Considered neutron-rich nuclei show the 
maximum skin thickness of the order of 0.9 fm.
The thickness particularly for the drip 
line nucleus is framed in Table ~\ref{tab2} and Table ~\ref{tab3}. 
Moreover, the skin thickness is also 
represented as a function of the asymmetry 
parameter ($\delta$=$N-Z/N+Z$) 
in Figs.~\ref{iskin1}, \ref{iskin2}. 
This graph shows the correlation of skin 
thickness with the asymmetry parameter. 
The trends of the set of data indicate an 
approximate linear dependency of neutron-skin 
thickness to the relative neutron excess of the nucleus.
It is demonstrated that our results match 
very well with available experimental
data~\cite{lubinski,friedman} within the error bars, 
and existing theoretical 
extractions~\cite{pickarewicz,hagen,mahzoon,rocamaza}.
Skin for $^{48}$Ca has been reported as 
$\Delta r=$ (0.176$\pm$0.018), (0.12-0.15), (0.249$\pm$0.023) 
with  density functional theory~\cite{pickarewicz}, 
ab initio calculations~\cite{hagen}, 
dispersive optical model~\cite{mahzoon}, respectively.
The present value of the skin for $^{48}$Ca is found 
to be 0.245 fm which satisfies the earlier 
predictions~\cite{pickarewicz,hagen,mahzoon,dhiman}.

It is evident from Table ~\ref{tab1} that there is no such 
difference within both interactions and 
NL3* is just the improved version of NL3 force. 
As we have already mentioned in the first section that 
the neutron skin thickness is strongly correlated with
density dependence of neutron symmetry energy at 
saturation~\cite{chen,Brown2000,Horowitz2001,Typel2001,Furnstahl2002,
Karataglidis2002,centelles,Agarwal,Mondal,Reinhard,Angeli}. 
Not only this, the slope of the symmetry energy which expected 
to be the cornerstone of drip line, masses, densities, and
skin thickness of neutron-rich nuclei~\cite{roca}.
Here, the both forces (NL3 and NL3*) have almost 
similar values of slope parameters i.e. 
$L=$ 118.6 MeV and $L=$ 122.63 MeV for
NL3 and NL3* respectively, and therefore both
forces produce almost identical drip lines and neutron skin thickness.

The predicted neutron skin for exotic nuclei might be 
used to simulate the symmetry energy for the neutron-rich matter.
The slope parameter shall be varied with variable skin thickness. 
However, there are uncertainties in neutron skin 
produced by different models and it is a model dependent quantity. 
The results on skin thickness may also be used 
for determining or extracting 
the isovector part from nucleon-nucleon interaction. 
The isospin behaviour of nucleon interaction 
exhibits more precisely from heavy nuclei such as $^{208}$Pb.
Such nuclei have a large inequality in neutron and proton number and therefore show the existence of symmetry energy. 
So, very next we shall look for the skin of largely unequal heavy nuclei. 
Further, the skin structure shall be viewed by investigating the density distributions. 
In the very next subsection, we shall look at the density profile for neutron skin candidates.

\subsection{Density profile}
The variance of densities with internucleon separation ($r$) gives the insight of nucleons 
distributions inside the nuclei from the center to the surface region.
The central part of density is meaningful for characterizing the bubble or semi-bubble type 
structure~\cite{grasso,khan,ikram}, whereas the tail part 
has signified as far as halo and skin structure is concerned.  
Here, we plot the densities for the designated neutron-rich nuclei, showing 
the anomalous behaviour of $r_n$ in comparison to the empirical trend. 
The isotopes, for example, C, N, O, Ni, Kr, Rb, Sr, Y, and Zr are used to plot the neutron density 
to look for the tails as given in figures~\ref{den1}-\ref{den4}.
We witness a tail is increasing for O, Ni, Kr, Rb, Sr, Y, and 
Zr in comparison to other nuclei with increasing neutron numbers.
This may be understood that the added neutrons redistribute the whole shell structure of 
the nucleus and as a result, extended density distribution is formed.
Neutron and proton densities for stable nuclides 
(e.g.$^{12}$C,$^{14}$N,$^{16}$O,$^{58}$Ni,$^{84}$Kr,
$^{85}$Rb, $^{88}$Sr,$^{89}$Y,$^{90}$Zr) are 
also plotted for the sake of references and comparisons.
Exotic nuclei of these isotopes have a long tail indicating the skin structure. 
It is evident from Figs.~\ref{den3} and \ref{den4} 
the fall of peripheral density at $r=$ 4 fm is seen 
for stable nuclei of  $^{12}$C, $^{14}$N, $^{16}$O but the peripheral region goes on 
increasing with the increase in the number of neutrons and the 
end of density at $r=$ 6 fm. 
In the same fashion, for medium mass nuclei Ni, Kr, Rb, Sr, Y, and Zr, the peripheral 
density ends at $r=$ 10 fm for neutron-rich 
exotic nuclei instead of $r=$ 7 fm for stable cases.
This gradual increase in the amount of peripheral region 
due to the excessive number of neutrons is 
responsible for a large amount of neutron skin thickness.
In general, the density profile shows a noteworthy decent expansion of neutron density 
when contrasted with proton density at the tail part of considered nuclei.
The extension is equal to or more than 2 fm in these cases, which supports the strong skin structure.

\section{Summary and conclusion}
The charge radii, neutron radii, and neutron skin thickness over 
a series of the isotopic chains of isotopes C to Zr ranging from their $\beta$ stability line to the two-neutron drip line have been explored.
Theoretical calculations are made from axially deformed solution of Lagrange equations. 
To understand the size of the isotopes, the charge radii are analyzed. 
Binding energy per particle is plotted to expose the maximum stability of the nuclides. 
To understand the skin structure, neutron radii are plotted as functions of neutron number and compared with 
empirical estimations ($r=r_0N^{1/3}$). Calculated theoretical results are contrasted 
with accessible experimental data as well as different theoretical predictions 
and a satisfactory agreement is observed. We also estimated 
the two-neutron separation energy by which drip line nuclei are marked and traditional magic numbers are seen. 
BCS pairing scheme has some limitation in drip line region and therefore for more 
convincing results a better pairing approximation is needed to care of continuum states.
To examine the skin structure more precisely, neutron density distribution is analyzed by plotting the density.  
The density profile shows a remarkable good extension of neutron density as compared 
to proton density at the tail part of considered nuclei.
Finally, neutron skin thickness is observed for the whole considered isotopic 
series and a maximum amount 1.053 fm is noticed for $^{134}$Y (NL3*). 
The separation energies, radii, and densities are considered to be the significant quantities
in order to examine the halo nuclei. In some cases, large diffused densities are encountered,
which may lead to halo structure in those exotic neutron-rich nuclei.
For example, $^{134}$Y is highly a diffused nucleus with extremely large $\Delta r$,
which implies that it may be a halo nucleus.
Further, cross-sectional studies on theoretical and experimental basis are needed to 
get a more clear picture of the existence of a halo structure.


\section*{Acknowledgements}
One of the author(UR) would like to thank University Grant Commission for providing UGC-SRF fellowship.


\end{document}